\documentclass[aip,jcp,preprint,nofootinbib,endfloats]{revtex4}
\usepackage{epsfig,epsf,subfigure,amsmath,amssymb}
\usepackage{color,ulem}
\usepackage{overpic}
\usepackage{graphicx}
\usepackage{dcolumn}
\usepackage{bm}
\usepackage{pslatex}

\begin{document}
\author{Santosh Mogurampelly$^{1}$}
%\email{santosh@physics.iisc.ernet.in}
\author{Swati Panigrahi$^{2}$}
\author{Dhananjay Bhattacharyya$^{2}$}
\author{A.~K.~Sood$^{3}$}
\author{Prabal K.~Maiti$^{1}$}\thanks{To whom correspondence should be addressed}
\email{maiti@physics.iisc.ernet.in}
\affiliation{
$^1$Centre for Condensed Matter Theory, Department of Physics, Indian Institute of Science,
Bangalore 560012, India \\
$^2$Biophysics Division, Saha Institute of Nuclear Physics, Kolkata 700064, India \\
$^3$ Department of Physics, Indian Institute of Science, Bangalore 560 012, India}
%\phone{+91 80 22932865}
%\fax{+91 80 23602602}
\title{Unraveling siRNA Unzipping Kinetics with Graphene}
%\keywords{Graphene, siRNA, Unzipping, Wrapping, Nanotube and van der Waals Interaction}

\begin{abstract}
Using all atom molecular dynamics simulations,
we report spontaneous unzipping and strong
binding of small interfering RNA (siRNA) on graphene. 
Our dispersion corrected density functional theory 
based calculations suggest that nucleosides of RNA have 
stronger attractive interactions with graphene as 
compared to DNA residues. These stronger interactions 
force the double stranded siRNA to spontaneously 
unzip and bind to the graphene surface. 
Unzipping always nucleates at one end of the 
siRNA and propagates to the other end after 
few base-pairs get unzipped. While 
both the ends get unzipped, the middle part 
remains in double stranded form because of 
torsional constraint. Unzipping probability 
distributions fitted to single exponential 
function give unzipping time ($\tau$) of 
the order of few nanoseconds which decrease 
exponentially with temperature. From the 
temperature variation of unzipping time 
we estimate the energy barrier to unzipping.
%These findings will have important consequences 
%while designing graphene based siRNA 
%delivery platforms and in graphene based
%nucleic acid sensors.
\end{abstract} 
\maketitle
\pagebreak
\section{Introduction}
The interaction of nucleic acids with carbon nanotubes (CNTs)
and graphene have attracted much attention due to the potential
applications in nanotube separation \cite{zheng1,zheng2},
sensing \cite{biosensor,mohanty}, sequencing 
\cite{garaj,schneider,christopher} and nanomedicine 
\cite{liu1,liu2,liu3,lu2010}~. Basic understanding of 
the interaction mechanism of nucleic acids with CNT or 
graphene \cite{anindyacpl,varghese_cpc,ralph2007,santoshjcp,santoshjbs,narahari} is 
essential for such applications.
%RNA interference (RNAi) is a powerful technology 
%for controlling the expression of genes in 
%biomedical applications. 
Small interfering ribonucleic acid (siRNA)
molecules are a class of double stranded non-coding RNA 
which are typically 21 to 23 nucleotides
in length. The properties of siRNA are 
actively being studied due to their potential 
influence on cell functionality and applications
in medicine to achieve RNA interference 
(RNAi) \cite{napoli,fire,couzin}~. 
The mechanism of RNAi involves RNA-induced 
silencing complex (RISC) comprising 
of Dicer, Argonaute2 and siRNA binding 
protein that induces unzipping of siRNA 
into two single strand RNAs \cite{zamore2000,
hutvagner2002,tomari2004,zamore2005,Ghildiyal2009}~.
One of these two strands acts as a guiding 
strand to form specific base-pairs with mRNA 
and silences the gene.
For using RNAi technology in the treatment
 of diseases such as cancer, HIV, viral infections and eye 
diseases \cite{kurreck}, efforts are being made for the efficient 
and safe siRNA delivery systems to achieve 
the desired RNAi effect. Dendrimers \cite{tsubouchi,
huang,vasumati} and carbon nanotubes \cite{liu1,liu2,liu3,santoshjcp}
are good carriers of siRNA into disease infected cell.
%%Various transfecting carriers such as linear 
%or branched cationic polymers
%(Dendrimer) \cite{tsubouchi,huang} , cationic 
%lipids \cite{elbashir,yang} , carbon
%nanotubes (CNTs) \cite{liu1,liu2,liu3} , cell 
%penetrating peptides \cite{morris,simeoni} and 
%few proteins \cite{puebla,minakuchi} can be used 
%for siRNA delivery.
The discovery of graphene has 
led to its possible use in efficient delivery 
of siRNA as well as various oligonucleotides.
Several attempts have been made to study the
properties of nucleic acid interaction with graphene
\cite{mohanty,varghese_cpc,ralph2007,zhao2011,narahari,lu2010,lv2010} .
However the interaction between graphene and siRNA/DNA 
has not been understood.\\

In this paper, we show the unzipping
of siRNA on graphene and subsequent binding
between the two. Graphene is a 
free-standing two dimensional monolayer of 
carbon atoms arranged into a honeycomb 
lattice \cite{novoselov2004,meyer2007}~.
Graphene is a partially hydrophobic molecule: 
its faces are highly hydrophobic while edges,
depending on functionalization, can be 
hydrophilic in nature \cite{swatijpcc2011}~. 
Thus it has the ability to pass through
hydrophobic lipid bilayer and can also interact with
 the hydrophilic head groups. Dispersion 
interaction including $\pi-\pi$ stacking serves as 
the major attractive interactions
 between the non-polar molecules \cite{paton}~.
The importance of dispersion interactions has been 
verified recently in analyzing the structure and 
energetics of the graphene-nucleobase complexes
\cite{swatijpcc2012} and in studying the unzipping of 
siRNA with single walled CNT \cite{santoshjcp}~.
Using state of the art all atom molecular 
dynamics simulation along with the {\it{ab-initio}}
quantum mechanical calculations, we give a 
comprehensive understanding of the structure 
and thermodynamics of the siRNA-graphene complex.
In the supplementary material \cite{supplementary}, we discuss the 
effect of force fields (FF) on the unzipping 
and adsorption of siRNA/dsDNA on graphene.\\
%To study the effect of curvature we have also 
%looked at the interaction of siRNA/DNA with 
%CNTs of various diameters.\\

%The mechanism of fluorescent light
%switching on and off by DNA-graphene complex sreves as a biosensor.
%Such DNA/RNA-graphene biosensors can be used to diagnose diseases
%like cancer, detecting toxicity of tainted food and detecting
%pathogens from biological weapons. Other tests also revealed
%that single/double-stranded DNA attached to graphene can not be easily
% degraded by enzymes and the structures remain stable, which
%phenomena could lead to drug delivery for gene therapy.
%The organization of the paper is as follows. The next section (section 2) 
%describes the computational methods and system preparation. Section 3 
%gives details of the results and discussion on siRNA wrapping due 
%to dispersion interaction, binding energy, thermodynamics of the 
%siRNA-CNT complex and number of close contacts of siRNA to CNT 
%calculations. Finally in the section 4 we give a brief summary 
%of the results and conclude.\\ 
\section{Computational Details}
\subsection{Classical simulations}
\begin{figure*}
        \centering
        \subfigure[]
        {
        \includegraphics[height=60mm]{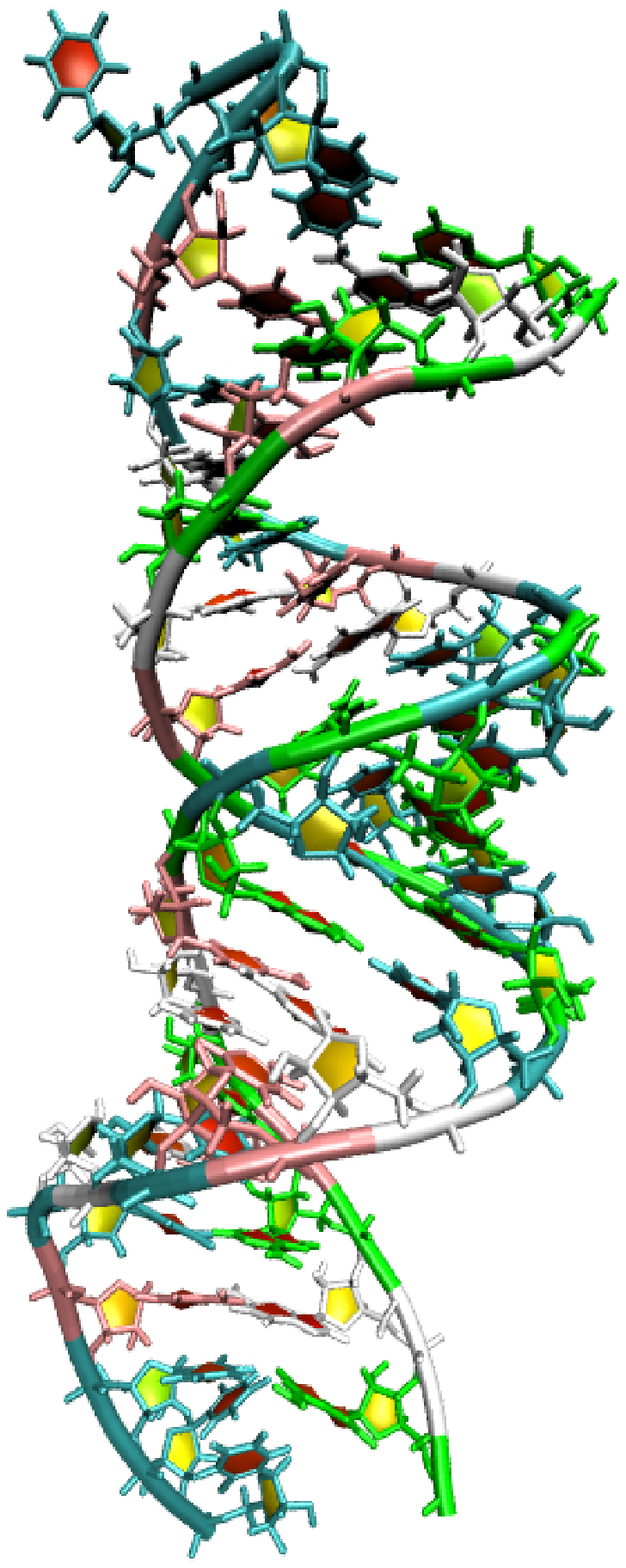}
        \label{sirna}
        }
        \subfigure[]
        {
        \includegraphics[height=60mm]{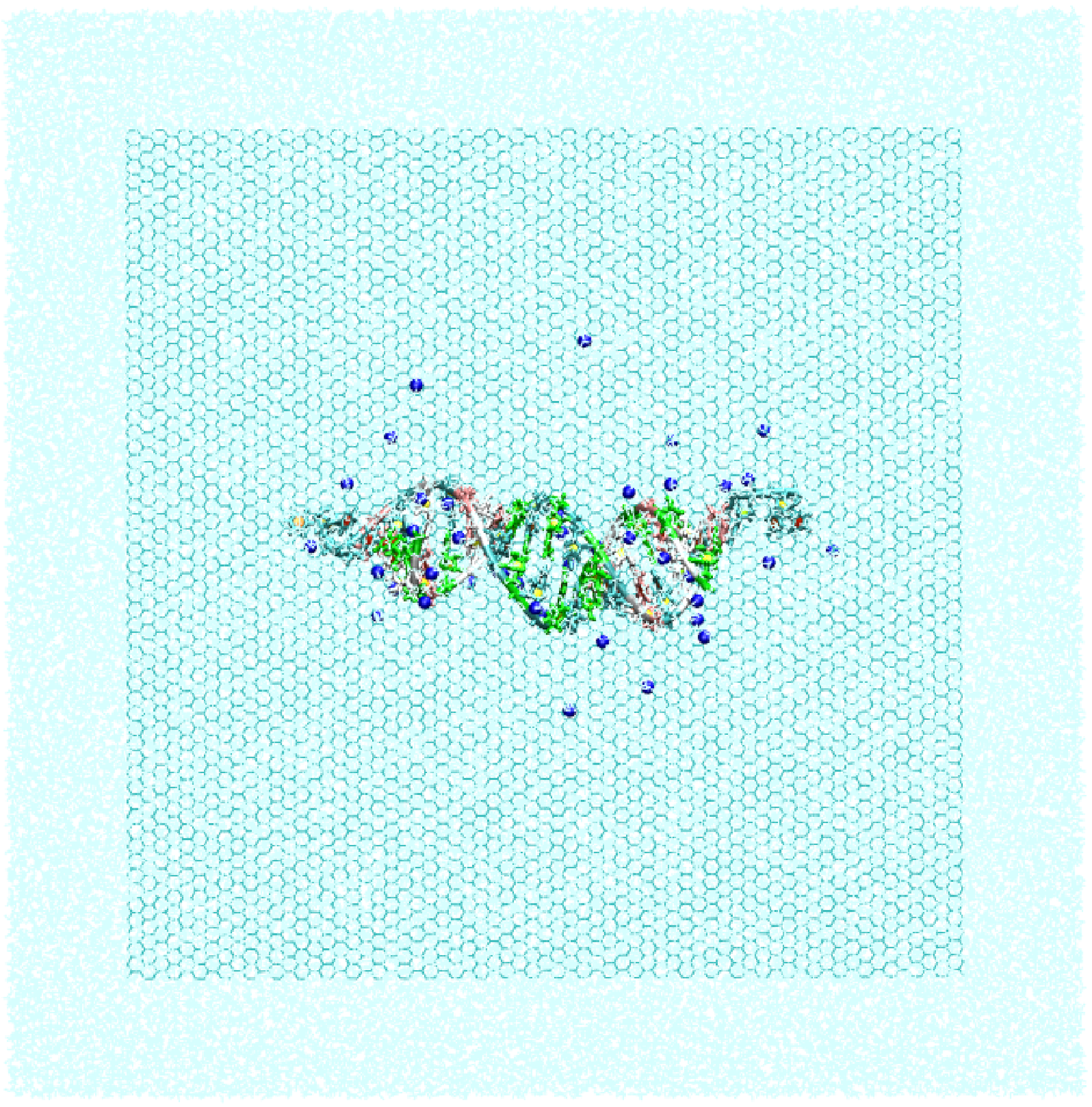}
        \label{sirna-ini}
        }
        \subfigure[]
        {
        \includegraphics[height=60mm]{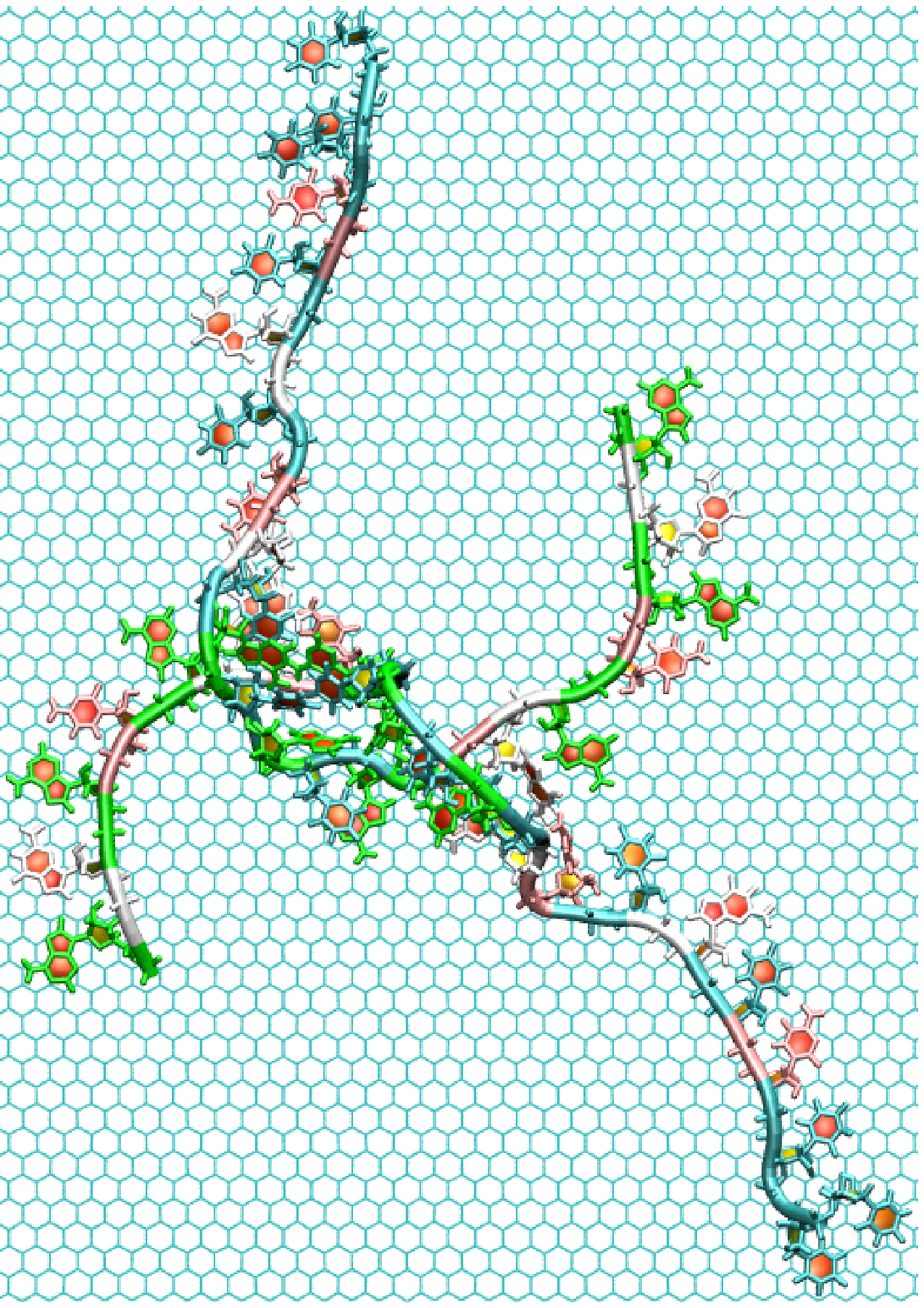}
        \label{40ns_view3}
        }
        \subfigure[]
        {
        \includegraphics[height=75mm]{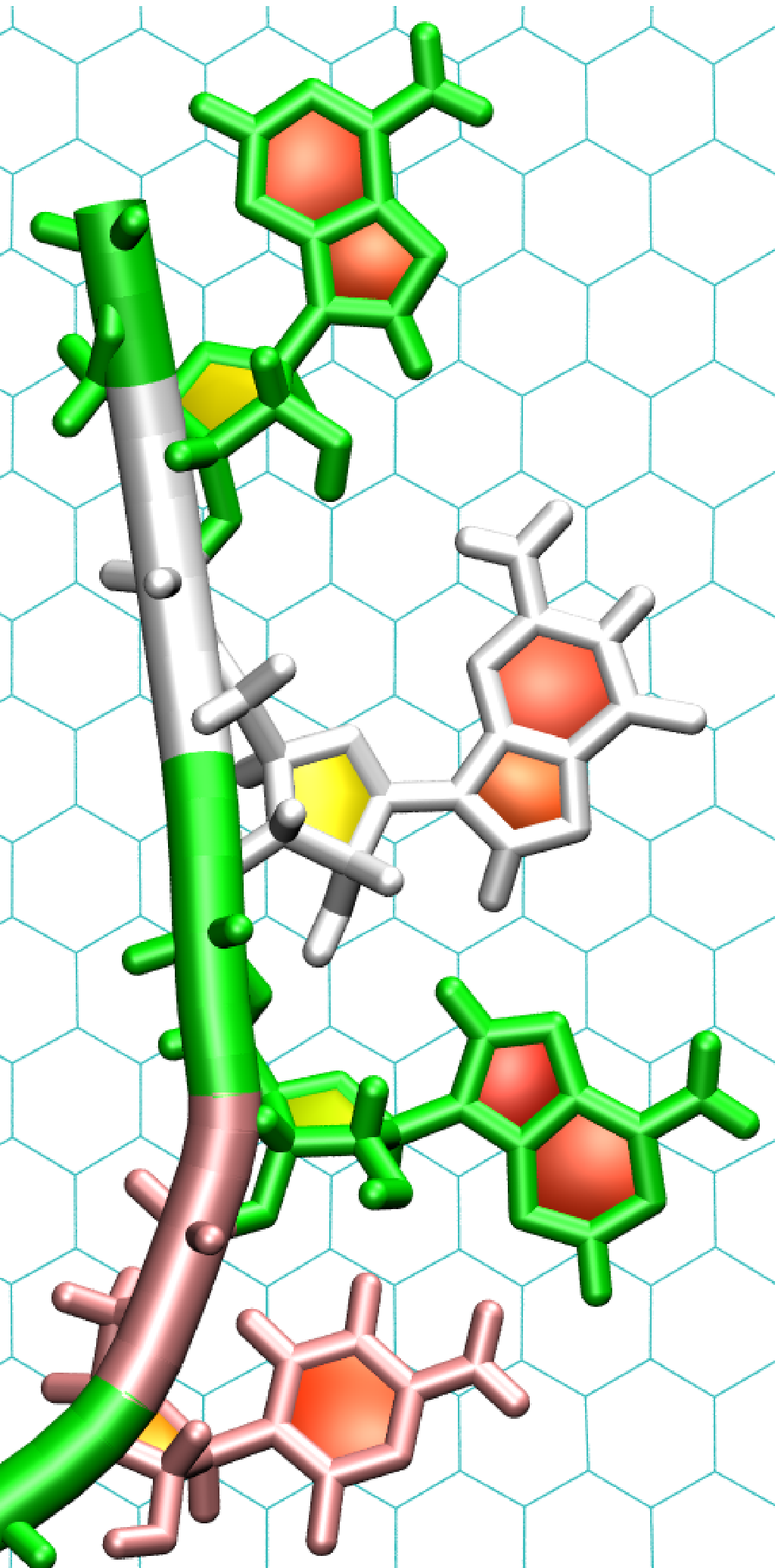}
        \label{40ns_view2}
        }
        \subfigure[]
        {
        \includegraphics[height=75mm]{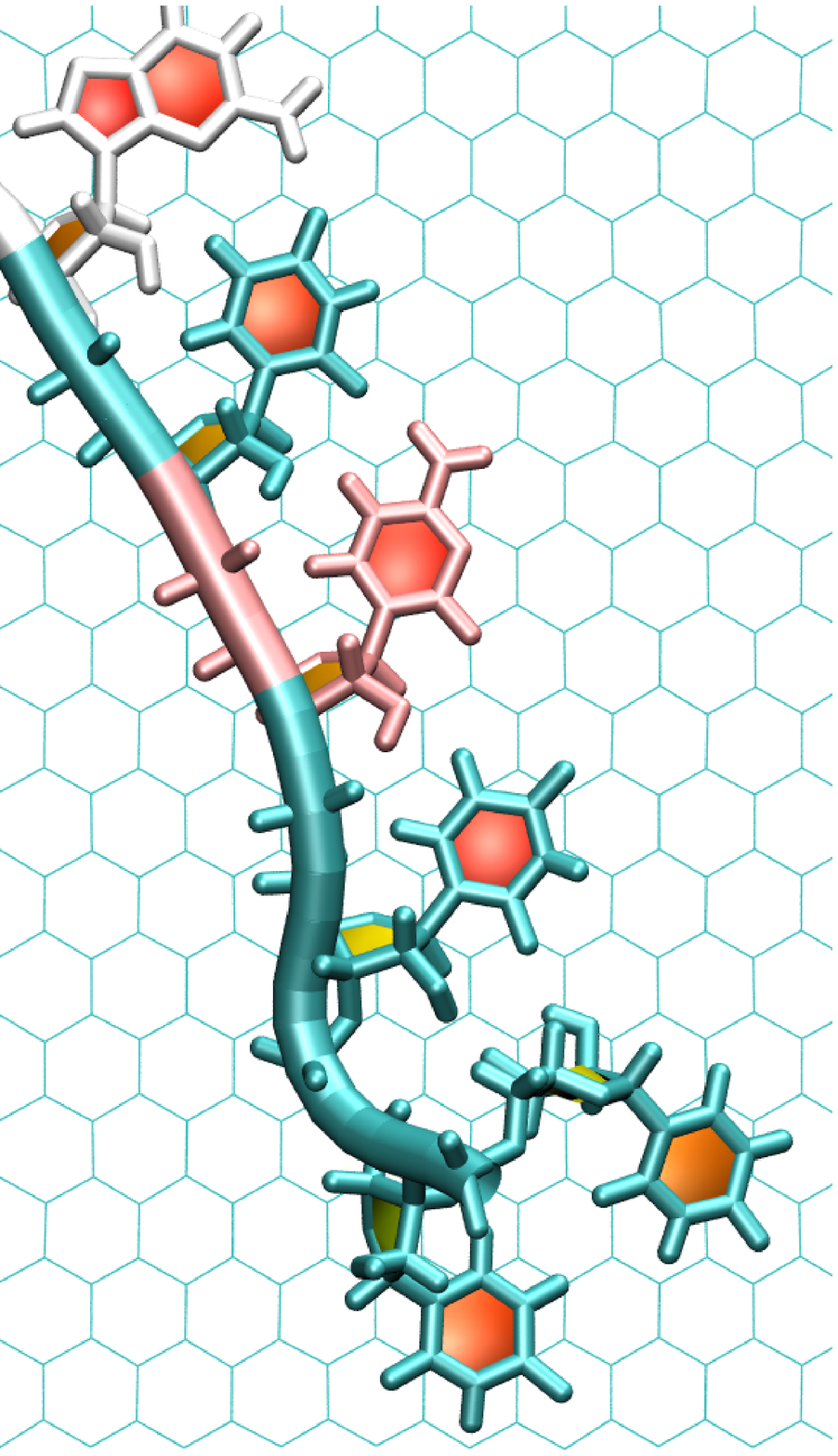}
        \label{40ns_view1}
        }
        \caption{(a) siRNA crystal structure (pdb code 1F8S) and
(b) the initial simulation system setup where siRNA-graphene complex
 was solvated with water and neutralizing Na$^{+}$ counterions.
(c) Snapshot of siRNA on graphene after optimum binding. (d) and (e)
shows the zoomed part of the completely unzipped siRNA strands whose
bases are in $\pi-\pi$ interaction with graphene.
The water and counterions were not shown in (c), (d) and (e) 
for image clarity. The snapshots were rendered using VMD software
package \cite{humphrey}~.}
        \label{snapshots}
\end{figure*}

We have used AMBER11 suite of programs \cite{amber11}
for simulating the systems with Amber 2003
 (along with ff99) force fields \cite{duan} and the TIP3P
model \cite{jorgensen} for water. Latest
improvements of torsion angle parameters 
are reported for DNA/RNA in a new force field called
parmbsc0 and ff10 \cite{parmbsc0,banas,yildirim}~.
The comparison of force fields ff99, parmbsc0 and ff10
are discussed in the supplementary material.
Similar conformational changes are observed with ff99,
parmbsc0 and ff10. 
%This gives us confidence
% that the reported results are not artefacts 
%of ff99 set of parameters.
Use of ff99 also helps us to compare 
the present results with our previous
simulations on siRNA-dendrimer complex
that used ff99 \cite{vasumati}~. 
The initial structure of
the small interfering ribonucleic 
acid (siRNA) was taken from the protein 
data bank (PDB code: 2F8S)\cite{yuan}~.
The sequence of the siRNA used is 
r(UU AGA CAG CAU AUA UGC UGU CU)$_{2}$ 
with sticky ends of sequence UU on the two
ends of the strands. We have built 
graphene sheet of $140\times 140~$ \AA$^2$~wider 
dimensions to ensure sufficient sliding
area for the siRNA before optimum binding on graphene.
For comparison, we have also simulated dsDNA of the
same length on graphene. The dsDNA has same sequence 
as the siRNA where the nucleobase Uracil (U) is 
replaced by Thymine (T) nucleobase, {\it{i.e.,}} d(TT AGA 
CAG CAT ATA TGC TGT CT)$_{2}$. Structure of the dsDNA 
in B-form was generated using NUCGEN module of AMBER \cite{amber11}~.
The siRNA/dsDNA-graphene
 complex structure is then solvated with TIP3P 
model water box using the LEaP module in AMBER 11 \cite{amber11}~.
The box dimensions were chosen such that there
is at least 20 \AA~solvation shell from the surface of
siRNA/dsDNA-graphene complex to the edge of the water box.
In addition, some water residues were replaced 
by 44 Na$^{+}$ counterions for siRNA and 42 Na$^{+}$ 
counterions for dsDNA to neutralize the negative
 charge on the phosphate backbone groups of the
siRNA/dsDNA structure. This has resulted in box
dimensions of 188$\times$186$\times$72 \AA$^3$~
with total system size of 214602 atoms for 
siRNA-graphene complex and 188$\times$186$\times$69 \AA$^3$~
with total system size of 206865 atoms for dsDNA-graphene 
complex. The crystal structure of 
siRNA and the initial system containing
siRNA and graphene with added water plus neutralizing
counterions are shown in Figures \ref{sirna} 
and \ref{sirna-ini}, respectively.\\

We have modeled the carbon atoms in graphene as 
uncharged Lennard-Jones particles with LJ 
parameters listed in Table \ref{lj}. 
%with a potential well
%depth of $\epsilon_C/k_B$ = 43.7 K, where $k_B$ is the 
%Boltzmann constant, and a diameter of $\sigma_C$ = 3.40 \AA.
%Similarly, the LJ parameters for Na+ are $\epsilon_{Na^+}/k_B$ 
%= 1.41 K and diameter is $\sigma_{Na^+}$ = 3.328 \AA and 
%for Cl- are $\epsilon_{Na^+}/k_B$ 
%= 134.7 K and diameter is $\sigma_{Cl^-}$ = 3.471 \AA.
The bonded interactions viz., stretching, torsion and 
dihedral terms were also included. To keep the 
graphene fixed during simulations, all the 
carbon atom positions in graphene were 
restrained with harmonic potential of spring 
constant of 1000 kcal/mol-\AA$^2$. The translational
 center of mass motions were removed every 1000 steps.
The system is subjected to standard simulation protocol
described in Refs. \cite{maiti2004,maiti2006}.
 The long range electrostatic interactions were calculated 
with the Particle Mesh Ewald (PME) method \cite{darden} 
using a cubic B-spline interpolation of order 4 and 
a $10^{-5}~$ tolerance is set for the direct space sum cutoff. 
A real space cutoff of 9 \AA~was used both for the long range
 electrostatic and short range van der Waals interactions with
 a non-bond list update frequency of 10. The trajectory was 
saved at a frequency of 2 ps for the entire simulation 
scale of 55-85 ns for each system.
We have used periodic boundary 
conditions in all three directions during the simulation. 
Bond lengths involving bonds to hydrogen
 atoms were constrained using SHAKE algorithm \cite{ryckaert}~. 
This constraint enabled us to use a time step of 2 fs for 
obtaining long trajectories of each 50 ns. During 
the minimization, the siRNA/dsDNA-graphene complex 
structures were fixed in their starting conformations
 using harmonic constraints with a force constant 
of 500 kcal/mol-\AA$^2$. This allowed the water 
molecules to reorganize which eliminates bad 
contacts with the siRNA/dsDNA and the graphene. 
The minimized structures were then subjected to 40 ps of MD,
 using 1 fs time step for integration. During the MD, the
 system was gradually heated from 0 to 300 K using 
weak 20 kcal/mol-\AA$^2~$ harmonic constraints on the 
solute to its starting structure. This allows slow 
relaxation of the siRNA/dsDNA-graphene complex structure.
Subsequently, simulations were performed under constant 
pressure-constant temperature conditions (NPT), with 
temperature regulation achieved
using the Berendsen weak coupling method \cite{berendsen} 
(0.5 ps time constant for heat bath coupling and 0.5 ps
pressure relaxation time). Constant temperature-pressure
MD was used to get the
correct solvent density corresponding to experimental
condition. Finally, for analysis of structures and 
properties, we have carried out 50 ns of NVT production 
MD with 2 fs integration time step using a heat bath 
coupling time constant of 1 ps.\\

The binding free energy for the non-covalent
association of two molecules in solution can be written as
$\Delta{G}\left(A+B\rightarrow{AB}\right)=G_{AB}-G_{A}-G_{B}$.
For any species on the right hand side $G(X)=H(X)-TS(X)$, accordingly,
\begin{equation}
\Delta{G}_{bind}=\Delta{H}_{bind}-T\Delta{S}_{bind}
\label{deltaG}
\end{equation}
where $\Delta{H}_{\text{bind}}=\Delta{E}_{\text{gas}}+\Delta{G}_{\text{sol}}.
~\text{Here}~\Delta{H}_{\text{bind}}$
is the change in enthalpy and is calculated by summing 
the gas-phase energies $\left(\Delta{E}_{\text{gas}}\right)$ 
and solvation free energies $\left(\Delta{G}_{\text{sol}}\right)$;
$E_{\text{gas}}=E_{\text{ele}}+E_{\text{vdw}}+E_{\text{int}}$,
where $E_{\text{ele}}$ is the electrostatic energy calculated
from the Coulomb potential, $E_{\text{vdw}}$ is the non-bond
van der Waals energy and $E_{\text{int}}$ is the internal
energy contribution from bonds, angles and torsions.
 $G_{\text{sol}}=G_{\text{es}}+G_{\text{nes}}$ where
 $G_{\text{es}}$ is the electrostatic energy calculated
from a Generalized Born (GB) method and $G_{\text{nes}}$
is the non-electrostatic energy calculated 
as $\gamma\times{SASA}+\beta$; where $\gamma$ is the 
surface tension parameter ($\gamma$ = 0.0072 kcal/mol-\AA$^2~$),
$SASA$ is the solvent-accessible surface area of the molecule
and $\beta$~is the solvation free energy for a point solute ($\beta$ = 0).
For the entropy calculation we have used two-phase 
thermodynamic (2PT) model \cite{lin2003,lin2010,hemant}~, based on density 
of states (DoS) function. The DoS function can be calculated 
from the Fourier transform of the velocity 
auto-correlation function which provides information 
on the normal mode distribution of the system.
The method has found successful application in 
several related problems 
\cite{lin2003,maitinl,lin2010,hemant,nandy2011}.

\subsection{Quantum simulations}
We have carried out quantum chemical calculations to
 understand the interaction of the planar
 nanographene with the nucleosides of RNA and DNA.
 For the present investigation, we have taken a graphene sheet of
 $(6\times6)$ dimension with alternate armchair and zig-zag
 edges with C-C bond lengths of 1.42 \AA,
 C-H {bond} length 1.09 \AA~ and all the angles kept at $120^\circ$. 
The edges are terminated with hydrogen atoms 
to avoid any unwanted terminal effects \cite{koskinen}~.
 We have modeled adenine, guanine, cytosine, thymine
 and uracil with furanose sugar connected to the respective 
nucleobases by the $\beta$-glycosidic bond. 
As structural features that distinguish siRNA
 from DNA are the presence of uracil nucleobase and 
hydroxyl-OH group in the constituted ribose sugar
 in siRNA, we have modeled thymine with the deoxyribose 
sugar and all other bases are modeled with ribose sugar to 
be consistent with siRNA structure. 
These structures act as miniature models 
of siRNA and DNA to understand the interaction with 
the planar graphene molecule. Initially the nucleosides
are placed vertically above, around 4 \AA~ from the center
of the planar hexagonal ring of the graphene and 
the nucleobases lie parallel to the graphene sheet. 
All the model building was done with the help of Discovery
 studio 2.0 \cite{studio} and Molden \cite{molden}
 software.\\

We have optimized all the graphene-nucleosides
complex systems to get the {energy} minimized configuration.
For all the quantum chemical calculations, we have 
employed dispersion corrected density functional
 approach using {$\omega$B97XD}/6-31G** basis set\cite{chai} 
using Gaussian 09 \cite{gaussian}~. The total interaction 
energy of each system has been calculated 
using $E_{int} = E(complex) - EX_o(isolated \mbox{ graphene})
 - EY_o(isolated \mbox{ nucleoside}) + $~BSSE, where BSSE 
represents basis set superposition error, arising 
from the overlapping of the atomic orbitals. The
BSSE corrections has been calculated using Boys 
and Bernardi function counterpoise 
method \cite{boysbernadi}~. We have also carried out 
frequency calculation on the optimized
geometry using the same method and basis set.
To analyze the degree of binding, we have also 
carried out the charge transfer analysis of 
the nucleosides and graphene sheet using the 
Natural Bond Orbital (NBO) approach \cite{reed1985,
reed1988,carpenter1988}~.\\

\section{Results and discussion}
\subsection{siRNA Unzipping}
\begin{figure*}
        \includegraphics[height=65mm]{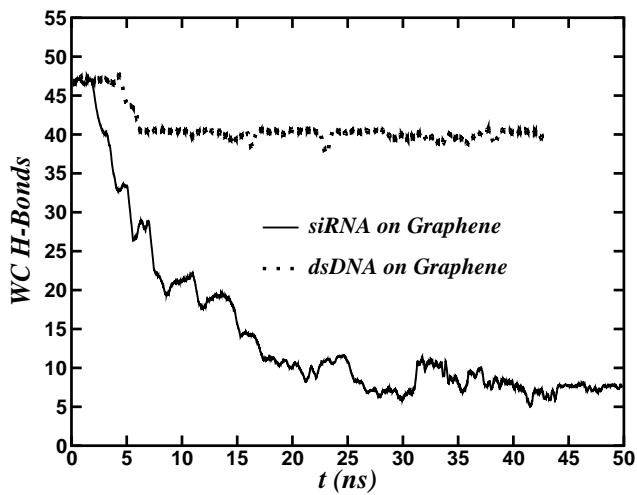}
        \caption{Number of intact Watson-Crick H-bonds in siRNA and 
dsDNA as a function of time with graphene. The siRNA has lesser intact WC H-bonds than dsDNA.}
        \label{wc}
\end{figure*}

Figure \ref{40ns_view3} shows the instantaneous snapshots of 
siRNA as it binds to graphene.
The unzipping of base-pairs starts
within 3 ns and by 22 ns, 41 of the initial 
48 Watson-Crick (WC) H-bonds are broken resulting 
in almost complete unzipping of siRNA into 
two single strand RNAs. Figure \ref{wc} shows 
the number of intact H-bonds of 
siRNA and dsDNA with time when bound to graphene.
Breaking of WC H-bonds in siRNA/dsDNA manifest the 
deformation mechanism of nucleic acid 
molecule \cite{voet,santoshjpcm,santoshbj}.
It can be seen that a large decrease of H-bonds
occurs in the first few nanoseconds when 
siRNA starts unzipping and within 22 ns, 41 H-bonds are broken 
leaving only 7 intact H-bonds. The unzipped 
bases are then free to interact with 
graphene surface via van der Waals 
forces. Figure \ref{wc} shows that H-bonds are saturated at 7.
At this stage, siRNA optimally 
binds to graphene and the complex is very stable 
after 22 ns.% which can be used as siRNA delivery 
%agent for RNAi applications.
In the optimum 
bound configuration where H-bonds are constant 
with time, there are few transient H-bonds due 
to room temperature thermal fluctuations. Because 
of these transient H-bonds, the curve has 
fluctuations about its mean value with standard 
deviations ranging from 1 to 3 H-bonds.
The different nucleosides interact with 
different strength with graphene making the binding
possible. Interestingly, dsDNA of same sequence 
and length show much less unzipping and 
binding on graphene compared to siRNA.
Within 6 ns, only 8 of total 48 H-bonds in dsDNA
get unzipped and remain constant at 40 H-bonds
throughout the rest of simulation time of 44 ns. This
difference has its origin in the extra
hydroxyl group of uridine that is present in siRNA.
To have a molecular level understanding of this 
binding affinity we calculate the binding free 
energy of different nucleosides with 
graphene both from the classical 
MD simulations (using MM-GBSA method) as 
well as from our dispersion corrected DFT calculations.
 
\subsection{Binding free energy}
\begin{figure*}
%        \subfigure[]
%        {
        \includegraphics[height=60mm]{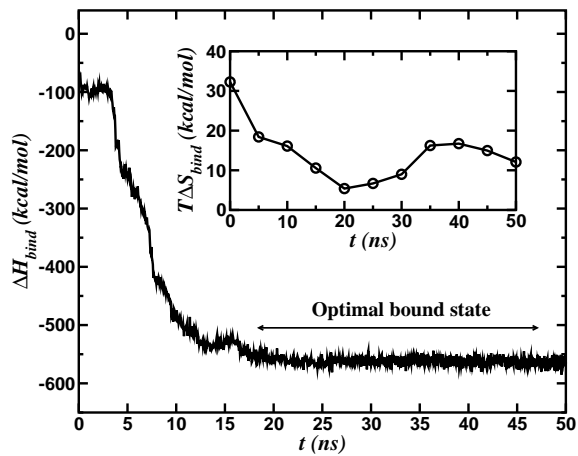}
%        \label{delta_gbtot}
%        }
%        \subfigure[]
%        {
%        \includegraphics[height=60mm]{free_energy_1bydiameter.eps}
%        \label{free_energy_diameter}
%        }
        \caption{Enthalpy of siRNA when binding to graphene as a 
function of time. Correspondingly, the entropy of siRNA is shown 
in the inset. The enthalpy gets saturated within 18 ns to -562.6 
kcal/mol. In the most optimum bound configuration, the binding 
free energy that includes enthalpy and entropy contributions 
is -573.0 $\pm$~ 8.0 kcal/mol.
}
        \label{delta_gbtot}
%        \label{Binding_free_energy}
\end{figure*}

Figure \ref{delta_gbtot} shows the enthalpy 
contribution to the total
binding free energy as a function of time as 
siRNA binds to graphene. In the plot, we have marked 
the time interval at which optimal binding 
happens as reflected by the binding energy. 
After this binding the complex is stable for 
the entire duration of the simulation
with fluctuations ranging only 1.5 \% of its average 
value in the optimum bound state. From the stable
trajectory of siRNA-graphene complex, 
the enthalpy contribution ($\Delta{H}_{bind}$) 
to the total binding free
energy was calculated for 250 snapshots 
separated each by 2 ps. 
The enthalpy contribution ($\Delta{H}_{bind}$) 
to the total binding free 
energy is -562.6 $\pm$ 6.2 kcal/mol.
Entropy is calculated
at every 5 ns along the trajectory. For this,
we simulate the system for 
40 ps with velocities and coordinates 
saved at a frequency of 4 fs.
The velocity auto-correlation function 
converges within 10 ps. When siRNA is
binding with graphene during initial stage, 
siRNA entropy decreases since 
the graphene substrate suppress the fluctuations 
of unzipped bases (inset of Figure \ref{delta_gbtot}).
 After siRNA binds optimally
to graphene, the entropy starts increasing 
again due to inherent kinetics of unzipped 
bases. However the entropy contribution ($T\Delta S_{bind}$) to 
the total binding free energy is small
compared to the enthalpy contribution ($\Delta{H_{bind}}$)
arising due to dispersive interaction between the 
graphene and the aromatic nucleobases. 
The enthalpy and entropy contributions
in Eqn. \ref{deltaG} give the total binding 
free energy ($\Delta G_{bind}$) of siRNA 
when binding to graphene. The value of $\Delta G_{bind}$
is -573.0 $\pm$ 8 kcal/mol. {However, the binding
energy of the dsDNA bound to graphene is calculated
to be -190 $\pm$ 9 kcal/mol which is very low
compared to siRNA bound to graphene.}\\

\begin{figure*}
        \subfigure[]
        {
        \begin{overpic}[height=61mm]{histogram_nucleoside.eps}
        \put(31,38){\includegraphics[scale=0.058]{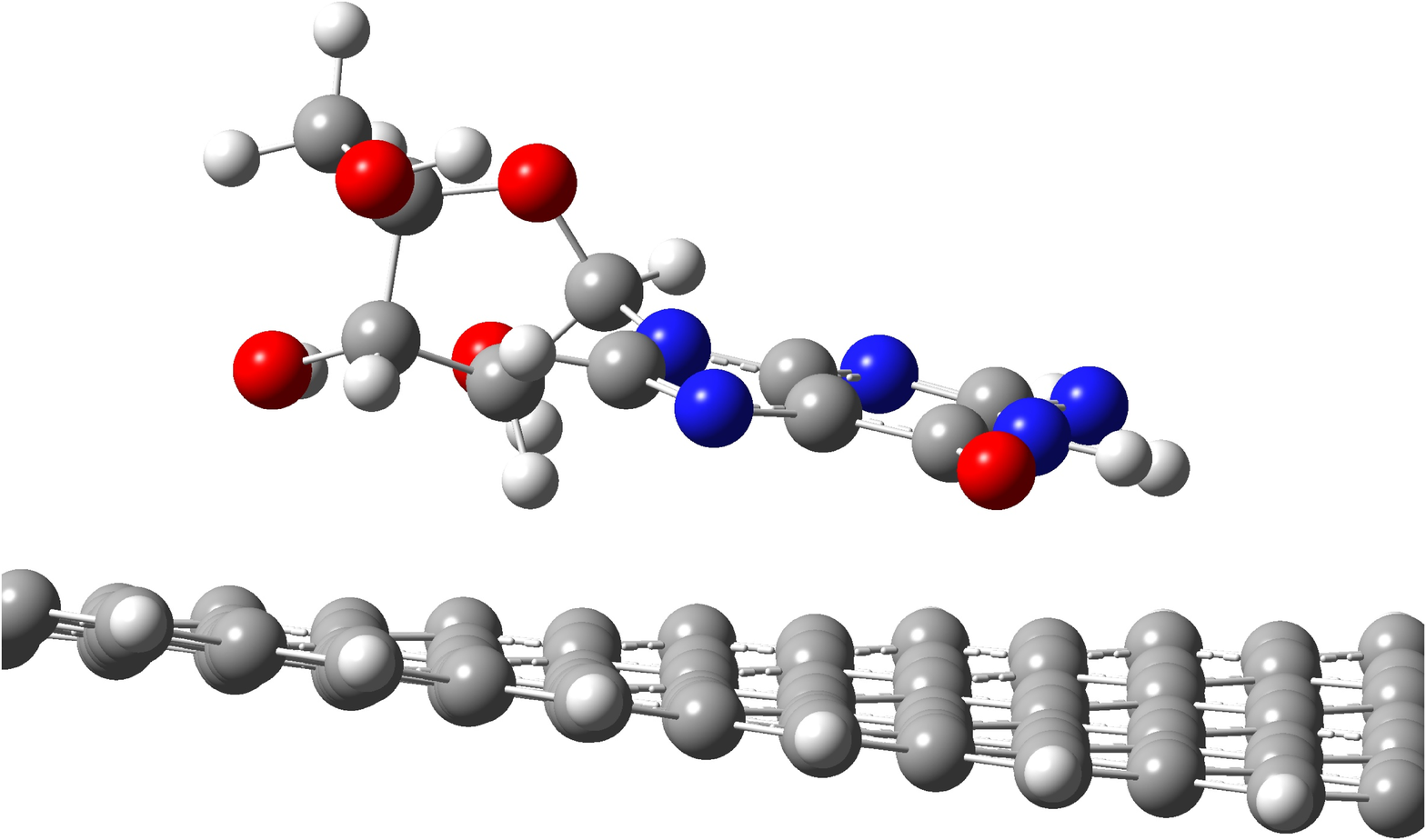}}
        \put(59,51){$\updownarrow$}
        \put(63,51){$r_\perp$}
        \end{overpic}
        \label{histogram}
        }
        \subfigure[]
        {
        \includegraphics[height=61mm]{free_energy_nucleoside.eps}
        \label{free_energy}
        }
        \caption{Nucleosides interaction with graphene from classical MD:
(a) Histogram of nucleosides position from graphene surface.
In the plot $r_{\perp}$ is the perpendicular distance between the center
of masses of nucleoside and graphene as shown. (b) Free
energy of the nucleoside-graphene complex as a function of $r_{\perp}$.
Minima in the free energy curve indicates the most optimum
bound configuration for the nucleoside-graphene complex. Inset
is the zoomed part around free energy minima. The minima occurs at a
value in the increasing order for guanosine, thymidine, adenosine, uridine and cytidine.
The smaller values of $r_{\perp}$ may imply a larger interaction strength
with graphene. Hence guanine has most interaction strength with graphene and
in decreasing order thymidine, adenosine, uridine and cytidine.}
        \label{histogram_free_energy}
\end{figure*}

To get further molecular level picture of the binding 
mechanism we also compute the histograms of the 
closest approach of nucleoside to graphene.
This will allow us to understand the relative 
binding affinity as well as how different 
nucleosides are oriented on the graphene substrate.
We track the center of mass position of different nucleobases
with respect to graphene as a function of simulation time.
We calculate the closest 
approach of nucleoside to graphene as the perpendicular
distance $r_{\perp} = \hat{z} \cdot \overrightarrow{AB} 
= B_z - A_z$ (because graphene is in $x-y$ plane) where $A$ and $B$ 
are the centers of mass position of nucleoside and graphene,
respectively. $r_{\perp}$ is calculated for 50 ns trajectory 
and the histogram of nucleosides is shown in 
Figure \ref{histogram}. From this plot, we can understand
the probability ($P(r_{\perp})$) of finding the 
nucleoside at a given position $r_{\perp}$
from the graphene. %\textcolor{red}{[You
%can also calculate how the nucleobases reparallel
%to the graphene sheet by calculating normals to each of
%the bases by NUPARM or by 3DNA].}
Using histogram, the free energy of nucleoside is calculated
as $F(r_{\perp}) = -k_BT~ln\left(P\left(r_{\perp}\right)\right)$
(where $k_B$ is the Boltzmann constant) and 
plotted as a function of distance in Figure \ref{free_energy}.
The minimum in the free energy indicates the most 
optimum bound configuration for nucleoside-graphene 
complex. In the inset we show the zoomed part of 
the occurrence of free energy minima. $F(r_{\perp})$ 
has minima at 3.775 \AA~, 3.790 \AA~, 3.795 \AA~,
3.840 \AA~ and 3.860 \AA~ for guanosine, Thymidine, adenosine,
uridine and cytidine, respectively. Guanine can 
form strongest H-bonding interaction
and stacking interaction with graphene.% \cite{}.
The distance of closest approach in the increasing 
order for guanosine, thymidine, adenosine, uridine and 
cytidine is $r_{\text{opt}}(G) < r_{\text{opt}}(T) \sim r_{\text{opt}}(A) 
< r_{\text{opt}}(U) < r_{\text{opt}}(C) $. 
Hence guanine has most interaction 
strength with graphene and cytidine has least 
interaction strength as G $>$ T $\sim$ A $>$ U $>$ C.
The binding energy order of nucleosides 
is {consistent} with experimental and 
theoretical calculations \cite{anindyacpl,varghese_cpc,ralph2007}~.
The stable complex structure with most of the 
 base-pairs already unzipped in siRNA can be 
delivered to the target
virus infected cell for RNAi applications. 
Since siRNA has to undergo unwinding process with the
effect of RISC, our proposed 
delivery mechanism by graphene possesses potential
advantages in achieving RNAi. Toxic effects of 
graphene inside cell may be suppressed with proper
surface functionalization \cite{lam,jia,cui}.\\

\subsection{Structural deformation}
\begin{figure*}
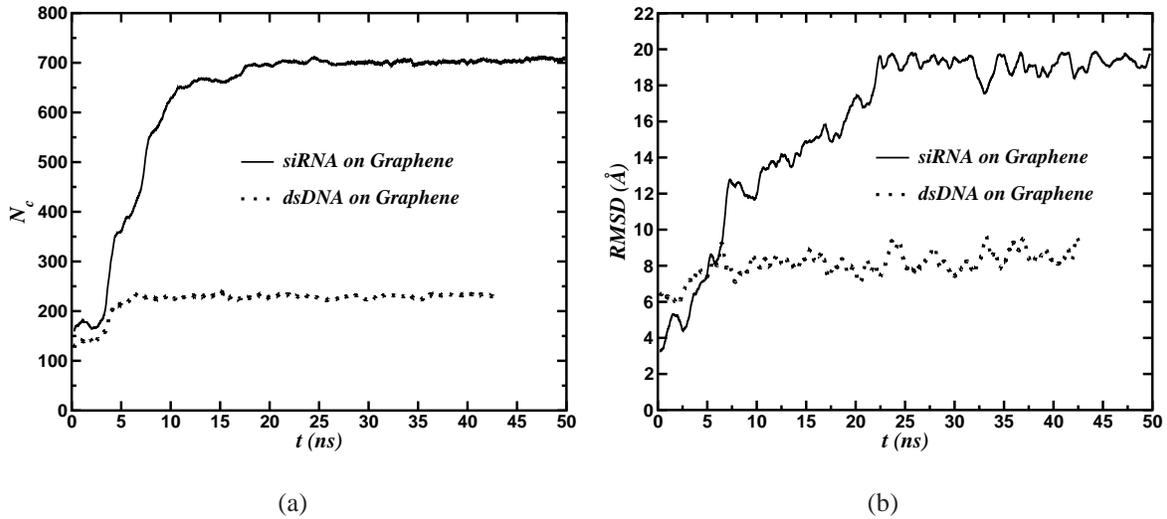

        \subfigure[]
        {
        \includegraphics[height=60mm]{cc_diameter_5A.eps}
        \label{cc} 
        }
        \subfigure[]
        {
        \includegraphics[height=60mm]{rmsd_rna_diameter.eps}
        \label{rmsd}
        }
        \caption{Number of the close contacts $N_c$ and RMSD of siRNA and dsDNA 
as a function of time. For siRNA, $N_c$ and RMSD are
maximum compared to dsDNA.}
        \label{cc_rmsd}
\end{figure*}

Snapshots shown in Figure \ref{snapshots} indicate that the siRNA 
molecule exhibit large structural deformation on
binding to graphene. This structural 
deformation is characterized by the number 
of siRNA atoms that come close to the graphene 
in a specified cutoff distance and the root mean
square deviation (RMSD) of siRNA with respect
to its crystal structure. We calculate close 
contacts $N_c$ when any siRNA atoms are 
within 5 \AA~of the graphene sheet. The number 
of close contacts $N_c$ between siRNA 
and graphene is plotted in Figure \ref{cc}
as a function of time. Since siRNA is getting 
unzipped within 21 ns, $N_c$ is increasing
rapidly within 21 ns and reaches a constant 
value of 710 siRNA atoms. The fluctuations are 
very less in $N_c$ after the complete binding 
of siRNA to the graphene. Interestingly, for 
dsDNA $N_c$ is much less than that of siRNA.
The dsDNA has only 240 atoms within 5 \AA~ from 
graphene sheet in the stable configuration.
This also demonstrates very low binding affinity 
of dsDNA compared to siRNA with graphene. In Figure \ref{rmsd} we 
plot the RMSD of siRNA/dsDNA as a function of time 
for both the siRNA/dsDNA-graphene complex.
For the calculation of RMSD, the reference 
structure of siRNA is taken to be the crystal 
structure of siRNA after initial minimization.
Note that the plot shows RMSD for only
production NVT simulation time scale. 
%Initially RMSD has 2 \AA~ in all the CNT 
%cases due to the structural deformation in 
%siRNA caused during 120 ps of NPT simulation.
In the most optimum bound 
configuration, the average RMSD 
$\left(\langle RMSD\rangle_{bound}\right)$
 of siRNA is 19.4 $\pm$ 0.4 \AA~ on graphene
whereas $\langle RMSD\rangle_{bound}$
of dsDNA is 8.5 $\pm$ 0.6 \AA. 
As the binding of siRNA is more on graphene, the siRNA 
structure deforms leading to a large 
value of $\langle RMSD\rangle_{bound}$.\\

\subsection{Unzipping kinetics}
\begin{figure*}
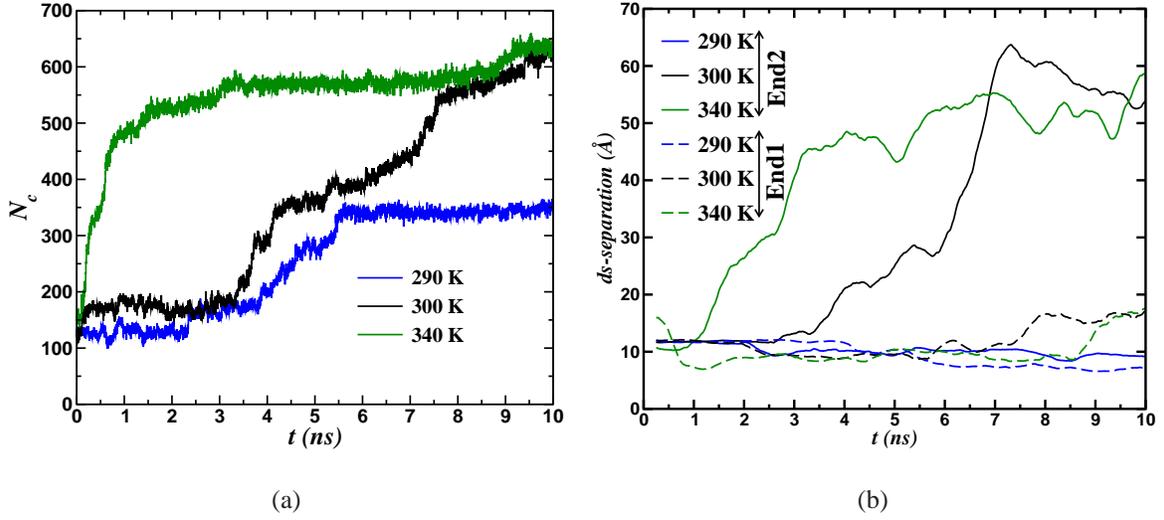

	\subfigure[]
	{
	\includegraphics[height=60mm]{cc_allraw.eps}
        \label{cc_temperature} 
	}
        \subfigure[]
        {
        \includegraphics[height=60mm]{dsseparation_temperature_end1_and_2.eps}
        \label{dssep_temperature}    
        }
\caption{Temperature effects: (a) Number of close contacts of 
siRNA on graphene (b) ds-separation of siRNA while unzipping 
from both the ends.}
\label{cc_dssep}
\end{figure*}

\begin{figure*}
\includegraphics[height=60mm]{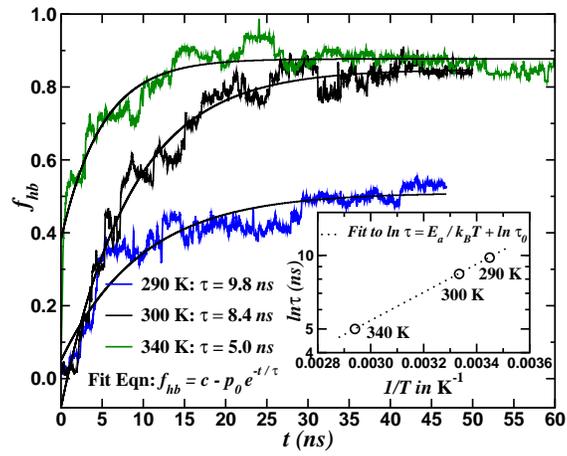}
\caption{Unzipping probability denoted by the fraction 
of broken WC H-bonds, $f_{hb}$ as a function of time for different 
temperatures.}
\label{tau_temperature}
\end{figure*}

In Figures \ref{cc_temperature} and \ref{dssep_temperature}, 
we plot the number of contacts 
for siRNA as well the distance between the two strands 
of the double stranded siRNA (ds-separation) at three 
different temperatures. It takes few ns to nucleate the 
unzipping. Once critical numbers of contacts are created 
between the siRNA and graphene, unzipping starts from 
one end (end2 for the current situation). Unzipped bases 
at one end help to make more contacts with graphene and 
thereby enhancing the interaction between siRNA and graphene.
This facilitates unzipping at the other 
end. Figure \ref{cc_temperature} shows that it takes almost 3 ns at 
300 K and less than 1 ns at 340 K for the critical 
number of contacts to be created. Once this is done, 
strong interaction between graphene and siRNA force 
the rapid unzipping of siRNA as is evident from the 
rapid increase of ds-separation in Figure \ref{dssep_temperature}.
To get an estimate of the unzipping time ($\tau$), we plot the 
unzipping probability distributions ($f_{hb}$) at three different 
temperatures in Figure \ref{tau_temperature} and fit them to single 
exponential functions as done in ref. \cite{mathe2004}.
This gives rise to $\tau$ = 9.8 ns, 8.4 ns and 5.0 ns 
at temperatures 290 K, 300 K and 340 K, respectively.
By fitting the unzipping time as a function of temperature
to $\tau = \tau_0 e^{E_a/k_BT}$, we get $\tau_0$ to
be 100.8 ps and the activation energy, $E_a$ to be 
2.637 kcal/mol or 0.114 $eV$. Plot of $ln\tau$ versus 
$\frac{1}{T}$ and the fit was shown in the inset 
of Figure \ref{tau_temperature}.\\

\subsection{Insights from Quantum simulations}
\begin{figure*}
	\subfigure[]
	{
	\includegraphics[height=40mm]{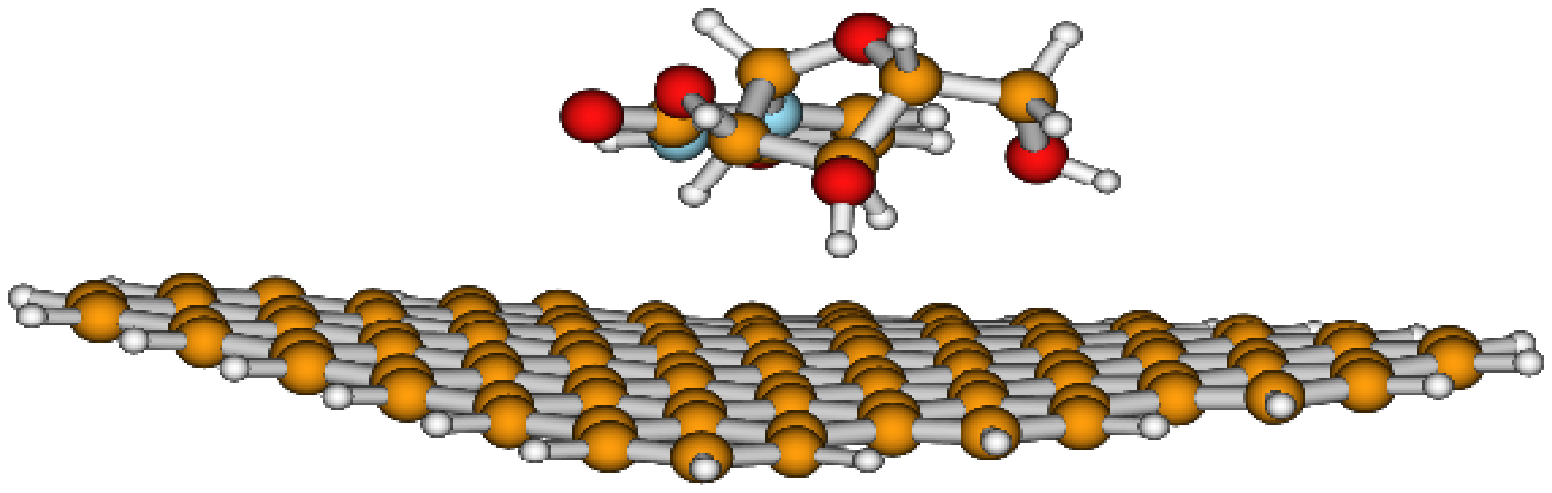}
        \label{ura} 
	}
        \subfigure[]
        {
        \includegraphics[height=40mm]{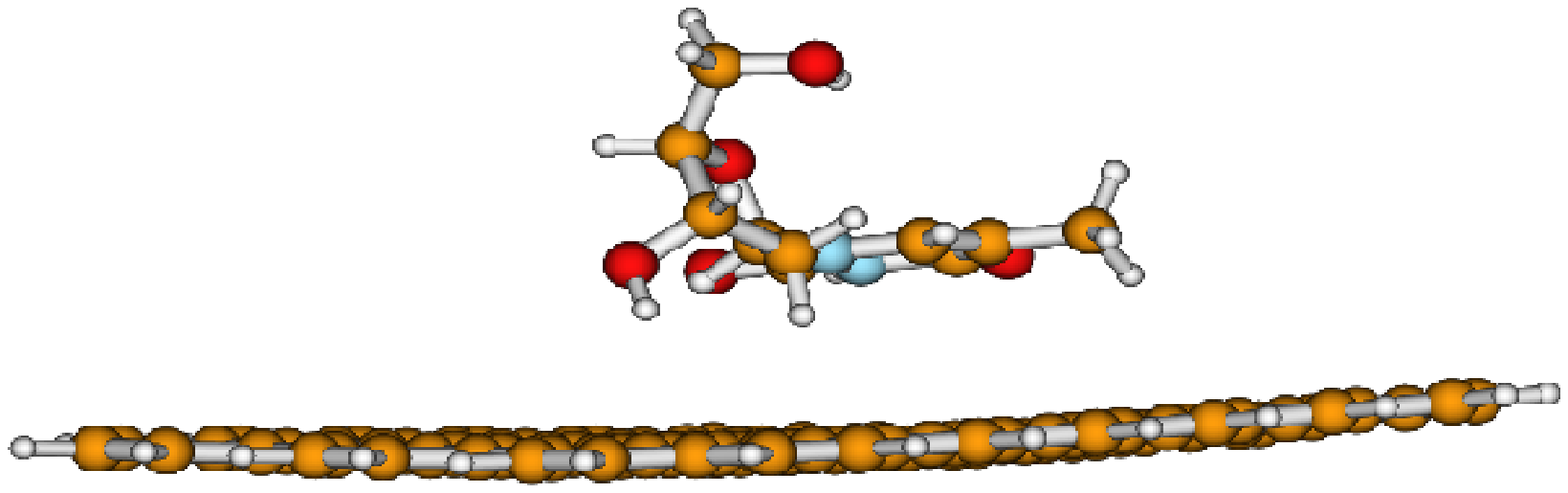}
        \label{thy}    
        }
\caption{Quantum DFT-D Optimized geometry of the complex systems: 
(a) graphene-uridine nucleoside and (b) graphene-thymidine nucleoside.}
\label{ura_thy}
\end{figure*}

To understand the difference in binding affinity
of siRNA and dsDNA with graphene we 
have calculated the binding energy of the 
graphene/siRNA and graphene/dsDNA miniature 
complexes using dispersion corrected DFT method (DFT-D).
{The sequence of the siRNA studied has A:U base pairs 
at both the ends, which is also the case
of most of the confirmed siRNA sequences \cite{chalk2005}, 
which are known to open up quite easily as compared to 
G:C base pairs. It may be noted that most of the DNA/RNA 
oligonucleotide sequences whose three-dimensional structures 
in double helical forms are available have G:C base pairs 
at both the termini \cite{dhananjay2009}. The siRNA 
sequences need to unzip soon for their functionality and 
probably that drives design of sequences with terminal 
loosely bound A:U base pairs. However, it appeared that 
the dsDNA molecule, containing terminal A:T base pairs 
does not unzip at physiological condition. In order to 
check whether this is due to stronger attraction in A:T 
as compared to A:U base pairs, we have optimized both
these base pairs using wB97XD/6-31G** method and found 
their interaction energies to be -15.80 and -15.91 kcal/mol, 
respectively. This clearly indicates that the terminal 
A:U base pair is not weaker one as compared to its DNA 
counterpart. Thus, the other component of differential 
interaction, i.e. interactions between Uracil residues 
and Thymine residues with graphene might be driving the 
RNA molecules to unzip.}
We are mainly interested in 
thymidine-graphene and uridine-graphene 
complex systems since these are the
 principal nucleosides, which can 
differentiate between DNA and RNA. 
Moreover these nucleosides remain
 unpaired in the siRNA as well as dsDNA. 
On analyzing the optimized geometry 
of the complex systems as shown in 
Figure \ref{ura_thy} (graphene-uridine and 
graphene-thymidine complex), we find 
that in both the cases 
O3$^{\prime}$-H3$^{\prime}$ of the 
constituent sugar points towards the 
planar graphene sheet with close approach 
forming O-H...$\pi$ contacts. In case of 
thymidine nucleoside, the closest 
O3$^{\prime}$-H3$^{\prime}$...ring center is found 
to be around 2.54 \AA~, and the angle 
$<$O3$^{\prime}$-H3$^{\prime}...$ring center$>$~ is 
obtained as $129.35^o$. These distances 
and angles are sufficient to form O-H...$\pi$ 
types of H-bonds. While in case of uridine
 nucleoside, the O3$^{\prime}$-H3$^{\prime}$...ring center
 bond distances are found to be 2.34 \AA~, and the angle 
$<$O3$^{\prime}$-H3$^{\prime}$...C$>$ is obtained 
as $165.32^{\circ}$, forming significantly stronger H-bond
between the graphene and uridine sugar,
as {compared} to that 
of thymidine sugar complex \cite{swatijmsd}~. 
The O2 and O4 groups of the thymine and uracil
 also interact with the graphene sheet, but the 
magnitudes of interaction seems to be very 
low as {compared} to that of O-H...$\pi$
 types of contact. Nevertheless these 
carbon oxygen atoms can form lone 
pair$\dots \pi$ type of contacts giving 
extra stabilization \cite{jain2009}~.
In addition to these contacts, 
the O2$^{\prime}$ may also form H-bond 
with the graphene, which however was not found in 
the energy minimized structures, but can not 
be ignored at physiological temperature
and between the oligonucleotides.
{Furthermore several of 2$^{\prime}$-OH groups, which are 
equivalent to 3$^{\prime}$-OH, are present in the
siRNA strands and absent in the dsDNA strands. 
Thus the siRNA strands tend to dissociate
from their double helical forms and bind 
strongly to the graphene sheet.}\\

The BSSE corrected interaction energy has been 
calculated for all the complex systems and
 they follows the trends G $>$ A $>$ U $>$ T $>$ C. 
Interaction energy of the graphene-uridine 
nucleoside is found to be -22.26 kcal/mol, 
while that of graphene-thymidine nucleoside 
is found to be -20.30 kcal/mol. So we can 
infer that uridine nucleoside interacts with
 the graphene more strongly than that of the 
thymidine nucleoside. These interaction energy
 values also well correlate the H-bond 
lengths and angles values obtained in both 
the cases. Our previous studies \cite{swatijpcc2012} 
on interactions of the graphene with the 
nucleobases gives the interaction energy 
strengths as G $>$ A $>$ C $>$ T $>$ U. So it 
proves that inclusion of sugar in the 
nucleobases alters the interaction energy 
strengths, therefore plays significant 
role in stabilizing the systems due to 
availability of more hydrogen bond donor 
sites.\\

Frequency calculation of the entire complex as well 
as the isolated systems give no imaginary frequency 
indicating the structures are at their 
local minima. Frequency calculation enables us to carry out 
thermochemical analysis of the system. All the 
calculations are carried out at 298.15 K and 1 atm.
 pressure. We have calculated the change in enthalpy 
and free energy of the systems. The $\Delta H$ of
graphene-uridine is found to be -26.39 
kcal/mol, whereas that of $\Delta H$ of
graphene-thymidine is calculated to be
-23.80 kcal/mol. Similarly the $\Delta G$ of 
graphene-uridine and graphene-thymidine 
are found to be -11.95 and -10.22 kcal/mol,
respectively. As it is well known that the system is
more favorable with increase in the negative value 
of $\Delta G$, graphene-uridine complex is more stable. 
The formation of stable graphene-uridine nucleoside 
complex may initiate and enhance the unzipping of 
the siRNA structure as observed by our counterpart 
MD simulation studies.\\

We have also calculated the NBO charges of the 
thymidine and uridine nucleosides of the graphene-nucleosides 
systems and compared with the isolated nucleosides. 
The difference in NBO 
charges of the major hydrogen bond donor atoms 
of the nucleosides which interacts with the 
graphene sheet are given in Table \ref{tabel2}.
We observed that charge transfer is more 
significant for the O3$^{\prime}$-H3$^{\prime}$
and O4 atoms for uridine molecule with graphene, 
since they interacts strongly with the graphene 
sheet. The O2 of thymidine shows negligible amount 
of charge transfer with the graphene sheet.
{It may also be noted that due to 
the methyl group of thymine, close to the O4 atom, 
the O4 atom of thymine may not be allowed to come close to the
graphene plane, thereby reducing its interaction strength.}
We can conclude that 
uridine interacts with the graphene sheet more 
strongly than thymidine, which is 
well correlated with the hydrogen bond strengths, 
interaction energy, and thermochemical analysis 
of the systems.\\
 
\section{Conclusion}
We demonstrated very unusual phenomena of complete 
siRNA unzipping and binding on graphene substrate.
{One of the major goals of the current study is to understand the binding mechanism of
siRNA/dsDNA on graphene. Our study also shows that siRNA unzips and binds to graphene
forming graphene-siRNA hybrid. This allows us to study the siRNA unzipping kinetics.
siRNA unzipping is very important in the context of RNAi therapeutics where a short siRNA
enters into cell and gets unzipped for its further action. Studying unzipping kinetics through
graphene also offers an alternate route to nanopore assisted unzipping where an electric field
is applied to translocate dsRNA/dsDNA. The stable graphene-siRNA hybrid may also be used
for efficient delivery of siRNA. The complex may penetrate the hydrophobic regions of the
bilayer due to hydrophobicity of graphene. Inside the cell, the graphene-siRNA may remain
as complex between graphene and two single stranded RNA chains. One of the chains might
easily dissociate from the complex whenever a competitive messenger RNA chain approaches
the complex. This can silence the required gene. Another interesting observation is that
dsDNA of same sequence as siRNA except thymine in place of uracil has less unzipping and
less binding on graphene. This interesting property could be used to detect or separate siRNA
and dsDNA molecules.}
We support these
findings through long classical MD 
simulation as well as calculating the 
relative binding affinity of nucleosides with 
graphene through dispersion corrected DFT methods.
It is shown that the unpaired uracil residues make 
strongest contacts with the graphene molecule 
through van der Waals and specific H-bonding 
interaction involving $2^{\prime}$-OH group of the 
ribose sugar. These interactions can be responsible
for the double helical siRNA to unzip, which are 
stabilized subsequently by several such O-H...${\pi}$ 
interaction. The equivalent double stranded DNA does 
not have the -OH group and hence remains stable throughout 
the simulation time. The spontaneous unzipping 
helps us to study the siRNA unzipping kinetics 
for the first time. Unzipping time is of the 
order of 5-10 ns and decreases with increasing 
temperature. Unzipping time follows Arrhenius 
behavior and allows us to get an estimate of 
the energy barriers for the siRNA unzipping. 
In contrast to the unzipping kinetics study 
through nanopore unzipping which requires 
application of voltage and takes longer 
time \cite{branton2003}, unzipping in 
graphene is very fast and happen 
spontaneously.\\

\section{Acknowledgements} 
We acknowledge computational resource supported 
by the DST Centre for Mathematical Biology at IISc. 
We thank DBT, India for the financial support. 
SM thank UGC, India for senior research fellowship.\\

%\listoftables
%\listoffigures
\clearpage

\begin{table}
\caption{Lennard-Jones interaction parameters used for 
carbon atoms in graphene and counterions Na$^{+}$.}
\begin{tabular}{*{3}{|c}|}
       \hline
        Atom          & $\epsilon/k_B$ (K) & $\sigma$ (\AA)\\
       \hline
        C        & 43.7  & 3.40  \\
        Na$^+$   & 1.41  & 3.328 \\
        \hline
        \end{tabular}
\label{lj}
\end{table}

\vspace{3 cm}
\begin{table}
\centering
\caption{Non bonded orbital (NBO) charges of the thymidine 
and uridine nucleosides}
\begin{tabular}{*{3}{|c}|}
       \hline
        Atom No. & Graphene + Thymidine & Graphene + Uridine\\
                 & difference in NBO charge & difference in NBO charge\\
       \hline
        O3$^{\prime}$-H3$^{\prime}$        & 0.005  & 0.007  \\
        O2   & 0.002  & 0.000 \\
        O4   & 0.007  & 0.014 \\
        \hline
        \end{tabular}
\label{tabel2}
\end{table}
\clearpage
\newpage

\clearpage
\clearpage

\clearpage

\clearpage

\clearpage

\clearpage

\clearpage

\clearpage

\begin{thebibliography}{99}
\bibitem{zheng1}
M.~Zheng, A.~Jagota, E.~D. Semke, B.~A. Diner, R.~S. Mclean, S.~R. Lustig,
  R.~E. Richardson, and N.~G. Tassi.
\newblock {\em {Nature Materials}}, {2}({5}):{338--342},  {(2003)}.

\bibitem{zheng2}
M.~Zheng, A.~Jagota, M.~Strano, A.~Santos, P.~Barone, S.~G. Chou, B.~A. Diner,
  M.~S. Dresselhaus, R.~S. McLean, G.~B. Onoa, G.~G. Samsonidze, E.~D. Semke,
  M.~Usrey, and D.~J. Walls.
\newblock {\em {Science}}, {302}({5650}):{1545--1548},  {(2003)}.

\bibitem{biosensor}
C.-H. Lu, H.-H. Yang, C.-L. Zhu, X.~Chen, and G.-N. Chen.
\newblock {\em {Angew. Chem. Int. Ed.}}, {48}({26}):{4785--4787}, {(2009)}.

\bibitem{mohanty}
N.~Mohanty and V.~Berry.
\newblock {\em {Nano Lett.}}, {8}({12}):{4469--4476},  {(2008)}.

\bibitem{garaj}
S.~Garaj, W.~Hubbard, A.~Reina, J.~Kong, D.~Branton, and J.~A. Golovchenko.
\newblock {\em {Nature}}, {467}({7312}):{190--193}, {(2010)}.

\bibitem{schneider}
G.~F. Schneider, S.~W. Kowalczyk, V.~E. Calado, G.~Pandraud, H.~W. Zandbergen,
  L.~M.~K. Vandersypen, and C.~Dekker.
\newblock {\em {Nano Lett.}}, {10}({8}):{3163--3167},  {(2010)}.

\bibitem{christopher}
C.~A. Merchant, K.~Healy, M.~Wanunu, V.~Ray, N.~Peterman, J.~Bartel, M.~D.
  Fischbein, K.~Venta, Z.~Luo, A.~T.~C. Johnson, and M.~Drndic.
\newblock {\em {Nano Lett.}}, {10}({8}):{2915--2921},  {(2010)}.

\bibitem{liu1}
Z.~Liu, M.~Winters, M.~Holodniy, and H.~Dai.
\newblock {\em {Angew. Chem. Int. Ed.}}, {46}({12}):{2023--2027}, {(2007)}.

\bibitem{liu2}
Z.~Liu, K.~Chen, C.~Davis, S.~Sherlock, Q.~Cao, X.~Chen, and H.~Dai.
\newblock {\em {Cancer Res.}}, {68}({16}):{6652--6660},{(2008)}.

\bibitem{liu3}
Z.~Liu, S.~Tabakman, K.~Welsher, and H.~Dai.
\newblock {\em {Nano Research}}, {2}({2}):{85--120}, {(2009)}.

\bibitem{lu2010}
C.-H. Lu, C.-L. Zhu, J.~Li, J.-J. Liu, X.~Chen, and H.-H. Yang.
\newblock {\em {Chem. Commun.}}, {46}({18}):{3116--3118}, {(2010)}.

\bibitem{anindyacpl}
A.~Das, A.~K. Sood, P.~K. Maiti, M.~Das, R.~Varadarajan, and C.~N.~R. Rao.
\newblock {\em Chem. Phys. Lett.}, {453}({4-6}):{266--273}, {(2008)}.

\bibitem{varghese_cpc}
N.~Varghese, U.~Mogera, A.~Govindaraj, A.~Das, P.~K. Maiti, A.~K. Sood, and
  C.~N.~R. Rao.
\newblock {\em ChemPhysChem}, {10}({1}):{206--210}, {(2009)}.

\bibitem{ralph2007}
{S.~Gowtham, R.~H.~Scheicher, R.~Ahuja, R.~Pandey, P.~Karna, and P.~ Shashi.
\newblock {\em Phys. Rev. B.}, {76}({3}):{033401}, {(2007)}.}

\bibitem{santoshjcp}
M.~Santosh, S.~Panigrahi, D.~Bhattacharyya, A.~K. Sood, and P.~K. Maiti.
\newblock {\em {J. Chem. Phys.}}, {136}({24}):{065106}, {(2012)}.

\bibitem{santoshjbs}
{B.~Nandy, M.~Santosh, and P.~K. Maiti.
\newblock {\em {J. Biosci.}}, {37}({3}):{457--474}, {(2012)}.}

\bibitem{narahari}
D.~Umadevi and G.~N. Sastry.
\newblock {\em J. Phys. Chem. Lett.}, 2(13):1572--1576, (2011).

\bibitem{napoli}
C.~Napoli, C.~Lemieux, and R.~Jorgensen.
\newblock {\em {Plant Cell}}, {2}({4}):{279--289},  (1990).

\bibitem{fire}
A.~Fire, S.~Q. Xu, M.~K. Montgomery, S.~A. Kostas, S.~E. Driver, and C.~C.
  Mello.
\newblock {\em {Nature}}, {391}({6669}):{806--811},  (1998).

\bibitem{couzin}
J.~Couzin.
\newblock {\em {Science}}, {298}({5602}):{2296--2297},{(2002)}.

\bibitem{zamore2000}
P.~D. Zamore, T.~Tuschl, P.~A. Sharp, and D.~P. Bartel.
\newblock {\em Cell}, 101(1):25 -- 33, (2000).

\bibitem{hutvagner2002}
G.~Hutvagner and P.~D. Zamore.
\newblock {\em Current Opinion in Genetics \& amp; Development}, 12(2):225 --
  232, (2002).

\bibitem{tomari2004}
Y.~Tomari, C.~Matranga, B.~Haley, N.~Martinez, and P.~D. Zamore.
\newblock {\em Science}, 306(5700):1377--1380, (2004).

\bibitem{zamore2005}
P.~D. Zamore and B.~Haley.
\newblock {\em Science}, 309(5740):1519--1524, (2005).

\bibitem{Ghildiyal2009}
M.~Ghildiyal and P.~D. Zamore.
\newblock {\em {Nature Rev. Genet.}}, {10}({2}):{94--108}, {(2009)}.

\bibitem{kurreck}
J.~Kurreck.
\newblock {\em {Angew. Chem. Int. Ed.}}, {48}({8}):{1378--1398}, {(2009)}.

\bibitem{tsubouchi}
A.~Tsubouchi, J.~Sakakura, R.~Yagi, Y.~Mazaki, E.~Schaefer, H.~Yano, and
  H.~Sabe.
\newblock {\em {J. Cell Biol.}}, {159}({4}):{673--683},  {(2002)}.

\bibitem{huang}
Y.~Z. Huang, M.~W. Zang, W.~C. Xiong, Z.~J. Luo, and L.~Mei.
\newblock {\em {J. Biol. Chem.}}, {278}({2}):{1108--1114}, {(2003)}.

\bibitem{vasumati}
V.~Vasumathi and P.~K. Maiti.
\newblock {\em {Macromolecules}}, {43}({19}):{8264--8274},  {(2010)}.

\bibitem{zhao2011}
X.~Zhao.
\newblock {\em {J. Phys. Chem. C}}, {115}({14}):{6181--6189},  {(2011)}.

\bibitem{lv2010}
W.~Lv, M.~Guo, M.-H. Liang, F.-M. Jin, L.~Cui, L.~Zhi, and Q.-H. Yang.
\newblock {\em {J. Mater. Chem.}}, {20}({32}):{6668--6673}, {(2010)}.

\bibitem{novoselov2004}
K.~Novoselov, A.~Geim, S.~Morozov, D.~Jiang, Y.~Zhang, S.~Dubonos,
  I.~Grigorieva, and A.~Firsov.
\newblock {\em {Science}}, {306}({5696}):{666--669},  {(2004)}.

\bibitem{meyer2007}
J.~C. Meyer, A.~K. Geim, M.~I. Katsnelson, K.~S. Novoselov, T.~J. Booth, and
  S.~Roth.
\newblock {\em {Nature}}, {446}({7131}):{60--63}, {(2007)}.

\bibitem{swatijpcc2011}
S.~Panigrahi, A.~Bhattacharya, D.~Bandyopadhyay, S.~J. Grabowski,
  D.~Bhattacharyya, and S.~Banerjee.
\newblock {\em {J. Phys. Chem. C}}, {115}({30}):{14819--14826},  {(2011)}.

\bibitem{paton}
R.~S. Paton and J.~M. Goodman.
\newblock {\em {J. Chem. inf. mod.}}, {49}({4}):{944--955},  {(2009)}.

\bibitem{swatijpcc2012}
S.~Panigrahi, A.~Bhattacharya, S.~Banerjee, and D.~Bhattacharyya.
\newblock {\em J. Phys. Chem. C}, 116(7):{4374--4379}, {(2012)}.

\bibitem{supplementary}
See Supplementary Material Document No. --- for figures, plots and discussion on the effect of force fields for 
ff99, parmbsc0 and ff10.

\bibitem{amber11}
D.~A. Case, T.~A. Darden, T.~E. Cheatham, III., C.~L. Simmerling, J.~Wang,
  R.~E. Duke, R.~Luo, R.~C. Walker, W.~Zhang, K.~M. Merz, B.~Roberts, B.~Wang,
  S.~Hayik, A.~Roitzberg, G.~Seabra, I.~Kolossvai, K.~F. Wong, F.~Paesani,
  J.~Vanicek, J.~Liu, X.~Wu, S.~R. Brozell, T.~Steinbrecher, H.~Gohlke, Q.~Cai,
  X.~Ye, J.~Wang, M.-J. Hsieh, G.~Cui, D.~R. Roe, D.~H. Mathews, M.~G. Seetin,
  C.~Sagui, V.~Babin, T.~Luchko, S.~Gusarov, A.~Kovalenko, and P.~A. Kollman.
\newblock Amber 11, (2010).

\bibitem{duan}
Y.~Duan, C.~Wu, S.~Chowdhury, M.~C. Lee, G.~M. Xiong, W.~Zhang, R.~Yang,
  P.~Cieplak, R.~Luo, T.~Lee, J.~Caldwell, J.~M. Wang, and P.~Kollman.
\newblock {\em {J. Comput. Chem.}}, {24}({16}):{1999--2012},  {(2003)}.

\bibitem{jorgensen}
{Jorgensen, W. L. and Chandrasekhar, J and Madura, J. D. and Impey, R. W. and
  Klein, M. L.}
\newblock {\em {J. Chem. Phys.}}, {79}({2}):{926--935}, {(1983)}.

\bibitem{parmbsc0}
A.~Perez, I.~Marchan, D.~Svozil, J.~Sponer, T.~E. Cheatham, III, C.~A.
  Laughton, and M.~Orozco.
\newblock {\em {Biophys. J.}}, {92}({11}):{3817--3829}, {(2007)}.

\bibitem{banas}
P.~Banas, D.~Hollas, M.~Zgarbova, P.~Jurecka, M.~Orozco, T.~E. Cheatham, III,
  J.~Sponer, and M.~Otyepka.
\newblock {\em {J. Chem. Theory Comput.}}, {6}({12}):{3836--3849}, 
  {(2010)}.

\bibitem{yildirim}
I.~Yildirim, H.~A. Stern, S.~D. Kennedy, J.~D. Tubbs, and D.~H. Turner.
\newblock {\em Journal of Chemical Theory and Computation}, 6(5):1520--1531,
  (2010).

\bibitem{yuan}
Y.-R. Yuan, Y.~Pei, H.-Y. Chen, T.~Tuschl, and D.~J. Patel.
\newblock {\em {Structure}}, {14}({10}):{1557--1565},  {(2006)}.

\bibitem{maiti2004}
P.~K. Maiti, T.~A. Pascal, N.~Vaidehi, and W.~A. Goddard.
\newblock {\em {Nuc. Acids Res.}}, {32}({20}):{6047--6056}, {(2004)}.

\bibitem{maiti2006}
P.~K. Maiti, T.~A. Pascal, N.~Vaidehi, J.~Heo, and W.~A. Goddard.
\newblock {\em {Biophys. J.}}, {90}({5}):{1463--1479}, {(2006)}.

\bibitem{darden}
T.~Darden, D.~York, and L.~Pedersen.
\newblock {\em {J. Chem. Phys.}}, {98}({12}):{10089--10092},{(1993)}.

\bibitem{ryckaert}
J.~P. Ryckaert, G.~Ciccotti, and H.~J.~C. Berendsen.
\newblock {\em {J. Comput. Phys.}}, {23}({3}):{327--341}, {(1977)}.

\bibitem{berendsen}
H.~J.~C. Berendsen, J.~P.~M. Postma, W.~F. Vangunsteren, A.~Dinola, and J.~R.
  Haak.
\newblock {\em {J. Chem. Phys.}}, {81}({8}):{3684--3690}, {(1984)}.

\bibitem{lin2003}
S.~T. Lin, M.~Blanco, and W.~A. Goddard.
\newblock {\em {J. Chem. Phys.}}, {119}({22}):{11792--11805}, {(2003)}.

\bibitem{maitinl}
P.~K. Maiti and B.~Bagchi
\newblock {\em {Nano Lett.}}, {6}:{2478--2485}, {(2006)}.

\bibitem{lin2010}
S.-T. Lin, P.~K. Maiti, and W.~A. Goddard, III.
\newblock {\em {J. Phys. Chem. B}}, {114}({24}):{8191--8198},{(2010)}.

\bibitem{hemant}
H.~Kumar, B.~Mukherjee, S.-T. Lin, C.~Dasgupta, A.~K. Sood, and P.~K. Maiti.
\newblock {\em {J. Chem. Phys.}}, {134}({12}):{124105},  {(2011)}.

\bibitem{nandy2011}
B.~Nandy and P.~K.~ Maiti.
\newblock {\em {J. Phys. Chem. B}}, {115}:{217--230}, {(2011)}.

\bibitem{koskinen}
P.~Koskinen, S.~Malola, and H.~Hakkinen.
\newblock {\em {Phys. Rev. Lett.}}, {101}({11}),115502, {(2008)}.

\bibitem{studio}
{\em {Discovery Studio 2.0, Accelerys Software Inc. San Diego, CA, USA}},
  {(2007)}.

\bibitem{molden}
G.~Schaftenaar and J.~Noordik.
\newblock {\em {J. Comput.-Aided Mol. Des.}}, {14}({2}):{123--134},
  {(2000)}.

\bibitem{chai}
J.-D. Chai and M.~Head-Gordon.
\newblock {\em {Phys. Chem. Chem. Phys.}}, {10}({44}):{6615--6620}, {(2008)}.

\bibitem{gaussian}
M.~J. Frisch, G.~W. Trucks, H.~B. Schlegel, G.~E. Scuseria, M.~A. Robb, J.~R.
  Cheeseman, G.~Scalmani, V.~Barone, B.~Mennucci, G.~A. Petersson,
  H.~Nakatsuji, M.~Caricato, X.~Li, H.~P. Hratchian, A.~F. Izmaylov, J.~Bloino,
  G.~Zheng, J.~L. Sonnenberg, M.~Hada, M.~Ehara, K.~Toyota, R.~Fukuda,
  J.~Hasegawa, M.~Ishida, T.~Nakajima, Y.~Honda, O.~Kitao, H.~Nakai, T.~Vreven,
  J.~A. Montgomery, {Jr.}, J.~E. Peralta, F.~Ogliaro, M.~Bearpark, J.~J. Heyd,
  E.~Brothers, K.~N. Kudin, V.~N. Staroverov, R.~Kobayashi, J.~Normand,
  K.~Raghavachari, A.~Rendell, J.~C. Burant, S.~S. Iyengar, J.~Tomasi,
  M.~Cossi, N.~Rega, J.~M. Millam, M.~Klene, J.~E. Knox, J.~B. Cross,
  V.~Bakken, C.~Adamo, J.~Jaramillo, R.~Gomperts, R.~E. Stratmann, O.~Yazyev,
  A.~J. Austin, R.~Cammi, C.~Pomelli, J.~W. Ochterski, R.~L. Martin,
  K.~Morokuma, V.~G. Zakrzewski, G.~A. Voth, P.~Salvador, J.~J. Dannenberg,
  S.~Dapprich, A.~D. Daniels, �.~Farkas, J.~B. Foresman, J.~V. Ortiz,
  J.~Cioslowski, and D.~J. Fox.
\newblock {\em Gaussian~09 {R}evision {A}.1}.
\newblock Gaussian Inc. Wallingford CT (2009).

\bibitem{boysbernadi}
S.~Boys and F.~Bernardi.
\newblock {\em Molecular Physics}, 19(4):553--566, (1970).

\bibitem{reed1988}
A.~E. Reed, L.~A. Curtiss, and F.~Weinhold.
\newblock {\em {Chem. Rev.}}, {88}({6}):{899--926}, {(1988)}.

\bibitem{reed1985}
A.~E. Reed, R.~B. Weinstock, and F.~Weinhold.
\newblock {\em {J. Chem. Phys.}}, {83}({2}):{735--746}, {(1985)}.

\bibitem{carpenter1988}
J.~E. Carpenter and F.~Weinhold.
\newblock {\em {J. Mol. Struc.-Theochem}}, {46}:{41--62},  {(1988)}.

\bibitem{voet}
D.~Voet and J.~G. Voet.
\newblock {\em Biochemistry}.
\newblock John Wiley \& Sons. Inc.,, 3 edition, (2005).

\bibitem{santoshjpcm}
M.~Santosh and P.~K. Maiti.
\newblock {\em {J. Phys.: Condens. Matter}}, {21}({3}):{034113},   {(2009)}.

\bibitem{santoshbj}
M.~Santosh and P.~K. Maiti.
\newblock {\em {Biophys. J.}}, {101}({6}):{1393--1402},{(2011)}.

\bibitem{lam}
C.-W. Lam, J.~T. James, R.~McCluskey, and R.~L. Hunter.
\newblock {\em Toxicological Sciences}, 77(1):126--134, (2004).

\bibitem{jia}
G.~Jia, H.~Wang, L.~Yan, X.~Wang, R.~Pei, T.~Yan, Y.~Zhao, and X.~Guo.
\newblock {\em Environmental Science \& Technology}, 39(5):1378--1383, (2005).

\bibitem{cui}
D.~Cui, F.~Tian, C.~S. Ozkan, M.~Wang, and H.~Gao.
\newblock {\em Toxicology Lett.}, 155(1):73 -- 85, (2005).

\bibitem{mathe2004}
J.~Mathe, H.~Visram, V.~Viasnoff, Y.~Rabin, and A.~Meller.
\newblock {\em {Biophys. J.}}, {87}({5}):{3205--3212},  {(2004)}.

\bibitem{chalk2005}
{A.~M. Chalk, R.~E. Warfinge, P.~Georgii-Hemming, and E.~E.~L. Sonnhammer.
\newblock {\em {Nuc. Acids Res.}}, {33}:{D131--D134}, {(2005)}.}

\bibitem{dhananjay2009}
{S.~ Samanta, S.~ Mukherjee, J.~ Chakrabarti, and D.~Bhattacharyya.
\newblock {\em {Nuc. Acids Res.}}, {130}:{115103}, {(2009)}.}

\bibitem{swatijmsd}
S.~Panigrahi, R.~Pal, and D.~Bhattacharyya.
\newblock {\em {J. Biomol. Struct. Dyn.}}, {29}({3}):{541--556}, {(2011)}.

\bibitem{jain2009}
A.~Jain, V.~Ramanathan, and R.~Sankararamakrishnan.
\newblock {\em {Protein Sci.}}, {18}({3}):{595--605}, {(2009)}.

\bibitem{branton2003}
A.~Sauer-Budge, J.~Nyamwanda, D.~Lubensky, and D.~Branton.
\newblock {\em {Phys. Rev. Lett.}}, {90}({23}):{238101},{(2003)}.

\bibitem{humphrey}
W.~Humphrey, A.~Dalke, and K.~Schulten.
\newblock {VMD: Visual molecular dynamics}.
\newblock {\em {J. Mol. Graph.}}, {14}({1}):{33--\&}, {(1996)}.

\end{thebibliography}
\end{document}

\author{Santosh Mogurampelly$^{1}$}
%\email{santosh@physics.iisc.ernet.in}
\author{Swati Panigrahi$^{2}$}
\author{Dhananjay Bhattacharyya$^{2}$}
\author{A.~K.~Sood$^{3}$}
\author{Prabal K.~Maiti$^{1}$}\thanks{To whom correspondence should be addressed}
\email{maiti@physics.iisc.ernet.in}
\affiliation{
$^1$Centre for Condensed Matter Theory, Department of Physics, Indian Institute of Science,
Bangalore 560012, India \\
$^2$Biophysics Division, Saha Institute of Nuclear Physics, Kolkata 700064, India \\
$^3$ Department of Physics, Indian Institute of Science, Bangalore 560 012, India}
%\phone{+91 80 22932865}
%\fax{+91 80 23602602}
\title{Supplementary material for\\ ``{\bf Unraveling siRNA Unzipping Kinetics with Graphene''}}
%\keywords{Graphene, siRNA, Unzipping, Wrapping, Nanotube and van der Waals Interaction}

\maketitle
\pagebreak
\section{Force field (FF) effects}
The results of siRNA unzipping mechanism 
on graphene discussed in the main paper are 
based on ff99 force field \cite{duan}~.
The ff99 force field has been widely used in the last 
decade to study the structure and thermodynamics 
of nucleic acids. However,
ff99 is found to produce unphysical torsion angle
transitions in `$\alpha/\gamma$' in longer MD simulations.
These irreversible `$\alpha/\gamma$' transitions
introduce severe distorsions in nucleic acids.
Based on these torsion angle parameterization artefacts,
we want to answer if the large structural transitions 
in siRNA/dsDNA when adsorbed to graphene are due to 
torsion angle artefacts of ff99.
Recently, torsion angle parameters for `$\alpha/\gamma$' 
and glycosidic dihedral `$\chi$' of nucleic acids have 
been refined in parmbsc0 \cite{parmbsc0,banas} 
and ff10 \cite{yildirim} force fields, respectively.
Note that ff10 is equivalent to ff99 + parmbsc0 for DNA
and ff99 + parmbsc0 + $\chi$ for RNA.\\

\begin{figure*}
        \subfigure[]
        {
        \begin{overpic}[height=60mm]{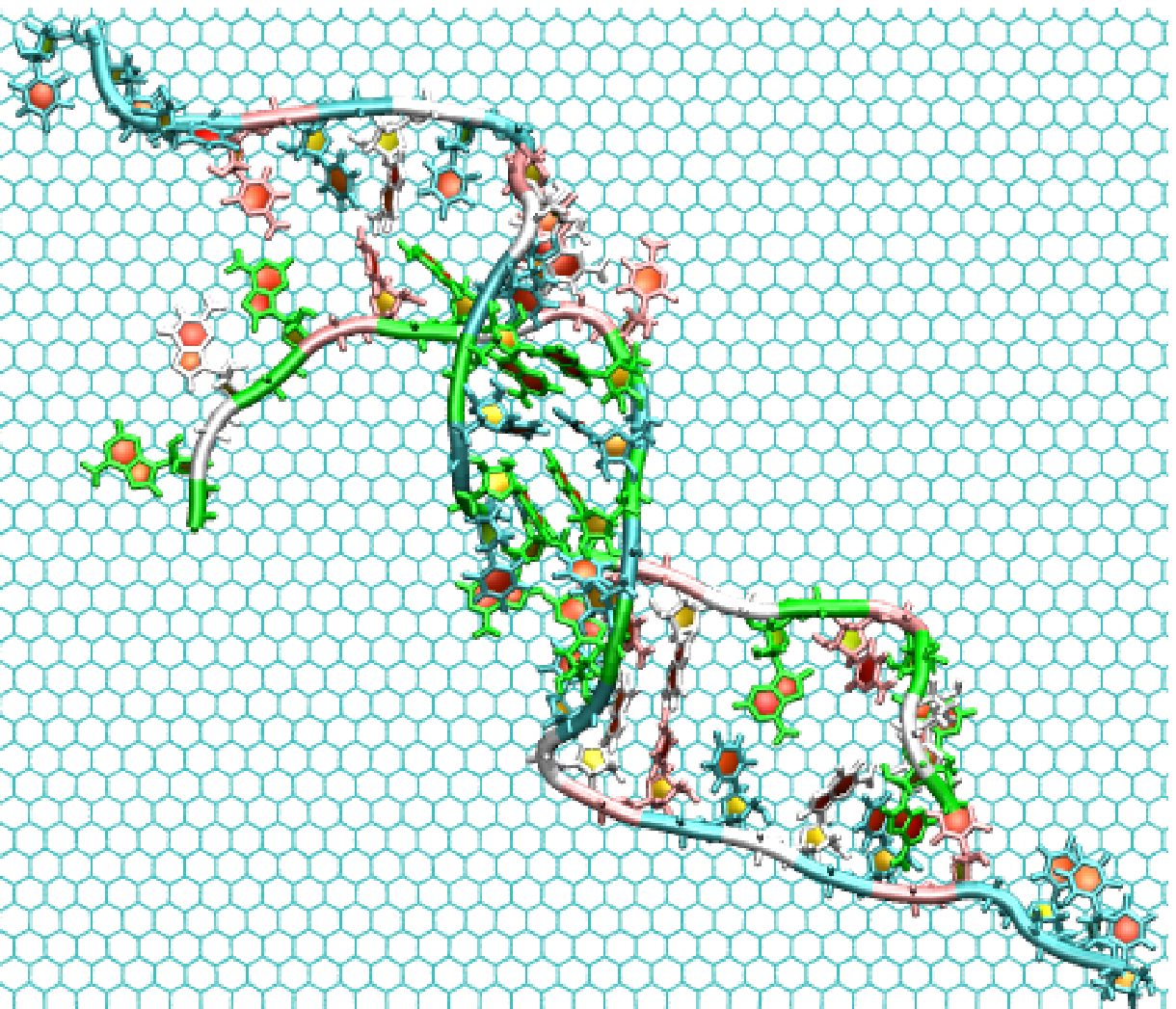}
        \put(5,5){siRNA with parmbsc0}
        \end{overpic}
%        \label{sirna_parmbsc0} 
        }
        \subfigure[]
        {
        \begin{overpic}[height=60mm]{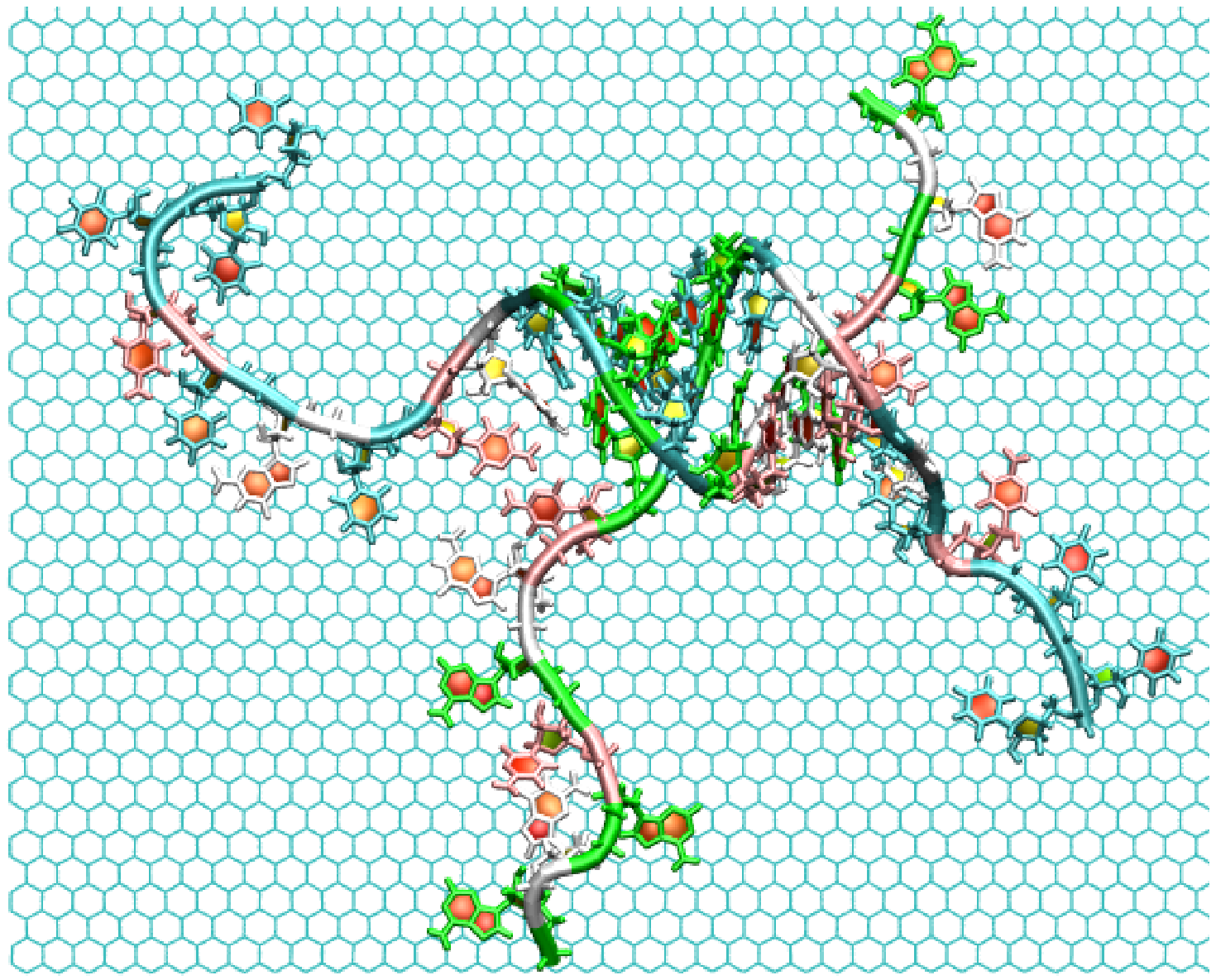}
        \put(56,5){siRNA with ff10}
        \end{overpic}
%        \label{sirna_parmbsc0} 
        }
        \subfigure[]
        {
        \begin{overpic}[height=60mm]{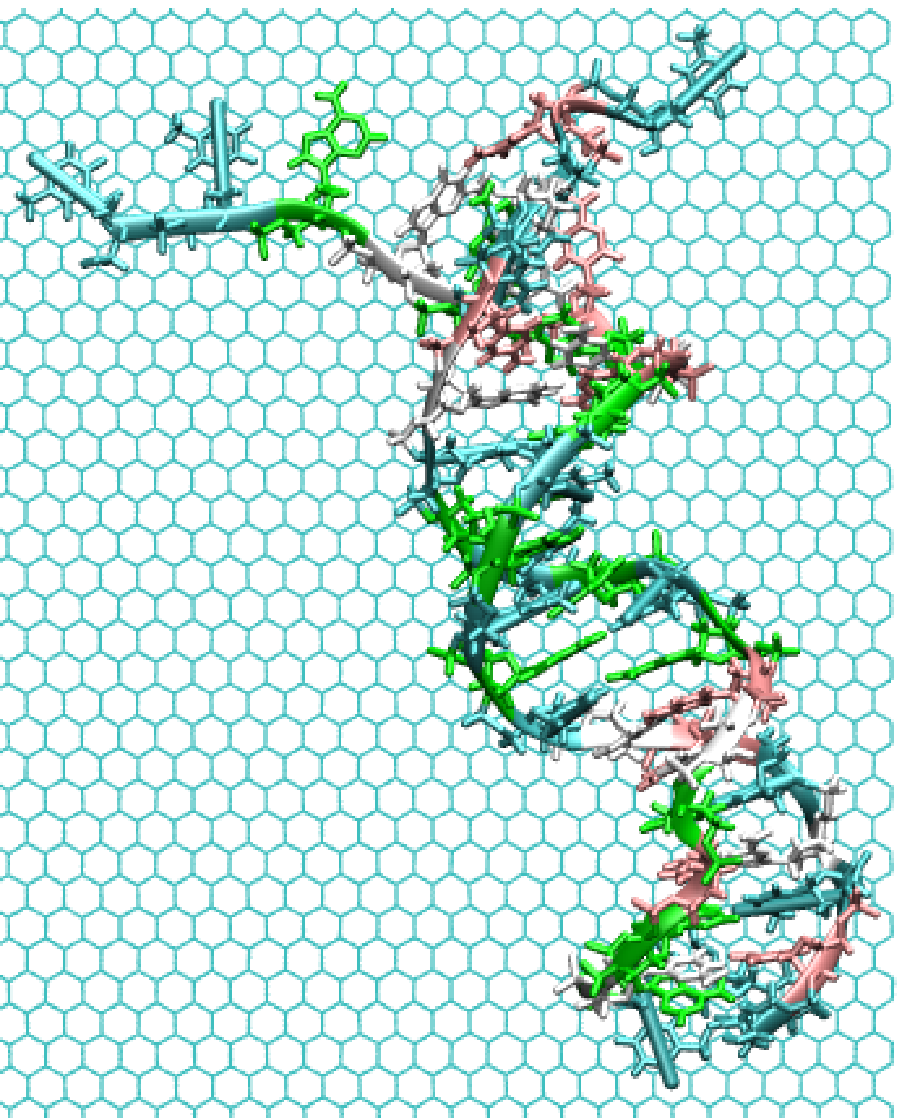}
        \put(5,5){dsDNA with ff99}
        \end{overpic}
%        \label{sirna_parmbsc0} 
        }
        \subfigure[]
        {
        \begin{overpic}[height=60mm]{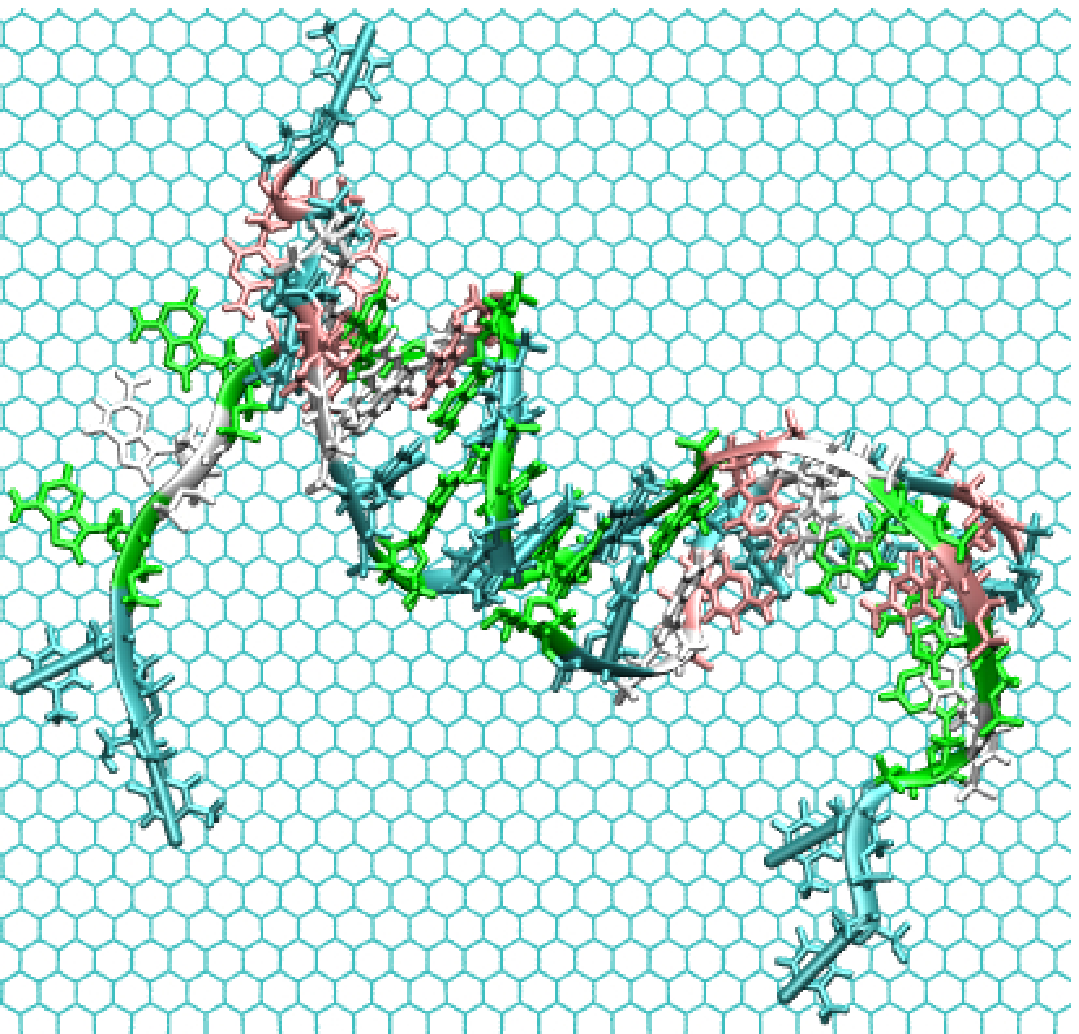}
        \put(5,5){dsDNA with parmbsc0}
        \end{overpic}
        \label{dsdna_parmbsc0}
        }
        \caption{FF effects: Snapshots of siRNA and dsDNA unzipping 
and adsorption on graphene using various force fields. The dimension 
of graphene sheet is 140$\times$140 \AA$^2$~ but only the portion of 
siRNA/dsDNA adsorption is shown in the figure. These snapshots are 
taken in the most optimum bound configuration both for siRNA and dsDNA.
With ff10 the unzipping of siRNA as can be quantified from the number 
of intact WC H-bonds (Figure \ref{wc_ff}) is less but unzipping 
and adsorption of siRNA on graphene is qualitatively similar to 
that observed with ff99/parmbsc0. However, parmbsc0 produces more unzipping 
of dsDNA than ff99 force field (Figure \ref{wc_ff}).}
        \label{ff_snapshots}
\end{figure*}

\begin{figure*}
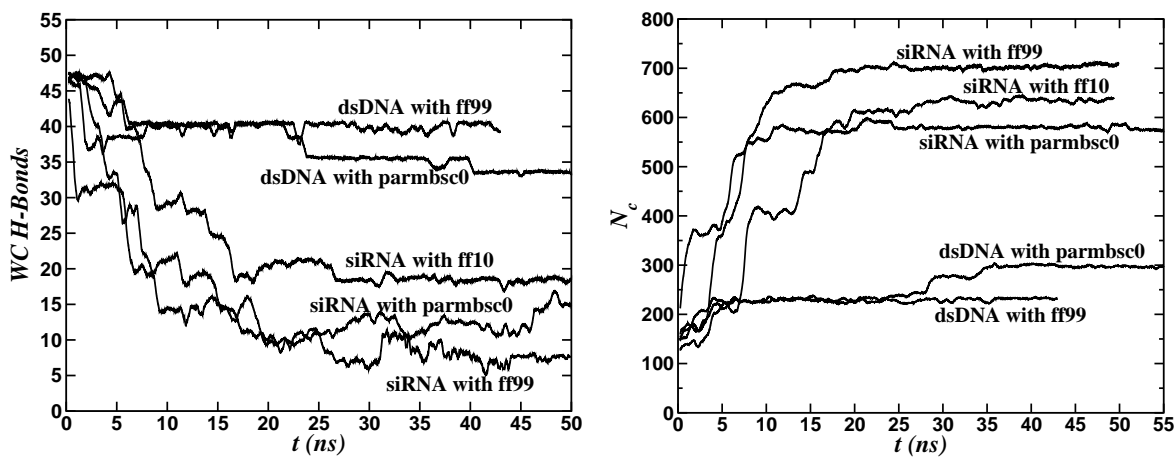
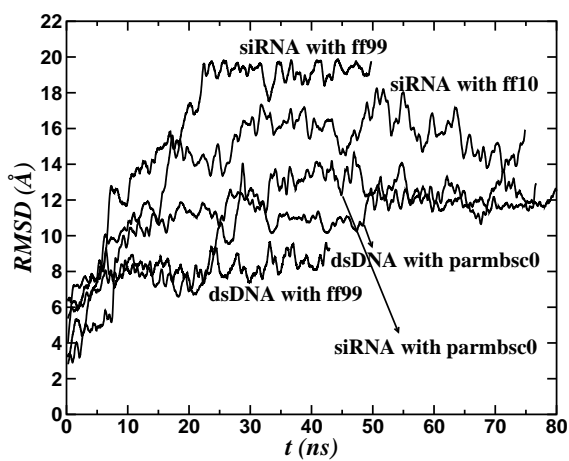

        \subfigure[]
        {
        \includegraphics[height=60mm]{wc_hbonds_ff99_parmbsc0_ff10.eps}
        \label{wc_ff} 
        }
        \subfigure[]
        {
        \includegraphics[height=60mm]{cc_5A_ff99_parmbsc0_ff10.eps}
        \label{cc_ff} 
        }
        \subfigure[]
        {
        \includegraphics[height=60mm]{rmsd_ff99_parmbsc0_ff10.eps}
        \label{rmsd_ff}
        }
        \caption{FF effects: Time series of the number of intact 
Watson-Crick H-bonds, close contacts and RMSD of
siRNA/dsDNA while adsorbing on graphene for various force fields.
These structural quantities verify that the unzipping and adsorption
of siRNA/dsDNA on graphene can also be observed using parmbsc0 and ff10.
It is interesting to note that RMSD is same for both the siRNA and dsDNA 
with parmbsc0 while they are very different when using ff99/ff10.}
        \label{ff_plots}
\end{figure*}

We have studied adsorption of siRNA/dsDNA on graphene with
ff99 and parmbsc0. In addition, siRNA adsorption on graphene
has been studied with ff10 also and as mentioned for dsDNA,
parmbsc0 is equivalent to using ff10. Snapshots and plots 
of siRNA/dsDNA on graphene with various force fields were
shown in Figures \ref{ff_snapshots} and \ref{ff_plots}, respectively. 
The major conclusion from these studies is that with 
parmbsc0 and ff10 also, unzipping and strong adsorption 
of siRNA on graphene is observed. In comparison,
dsDNA has less unzipping and adsorption on graphene 
compared to the unzipping and adsorption of siRNA.
Some of the earlier studies have also demonstrated 
that ff99 parameters are as good as parmbsc0 set in 
describing the conformational richness of 
RNA \cite{mccammon,bessova,reblova}~.
In our simulation with ff10, less unzipping of 
siRNA is observed compared to ff99/parmbsc0 (Figure \ref{wc_ff}).
In contrast, with parmbsc0, dsDNA has more unzipping and adsorption 
than ff99 as it is clear from snapshots (Figure \ref{ff_snapshots})
and WC H-bonds, close contacts and RMSD (Figure \ref{ff_plots}).
In spite of these differences, the qualitative features 
of the siRNA unzipping and adsorption 
on graphene have been reproduced with all the force fields 
under consideration. Hence, we conclude that the large
structural changes observed with ff99 force field are not
because of the artefacts of torsion angle parameters.\\
%
%\listoffigures
\clearpage

\clearpage

